\documentclass[final,10pt,3p,twocolumn]{elsarticle}

\journal{Journal of Magnetism and Magnetic Materials}

\usepackage[utf8]{inputenc}
\usepackage[T1]{fontenc}
\usepackage{amsmath}
\usepackage{amssymb}
\usepackage{graphicx}
\usepackage[skip=0pt]{subcaption}
\usepackage[format=plain,labelfont=bf]{caption}
\usepackage[version=3]{mhchem}
\usepackage[per-mode=symbol,range-phrase=--,range-units=single]{siunitx}
\usepackage{booktabs}
\usepackage{hyperref}
\usepackage{cleveref}
\usepackage{balance}
\usepackage{xr}
\usepackage{todonotes}
\presetkeys{todonotes}{inline}{}
\hypersetup{
  colorlinks=true,
}

\makeatletter
\providecommand{\doi}[1]{%
  \begingroup
    \let\bibinfo\@secondoftwo
    \urlstyle{rm}%
    \href{http://dx.doi.org/#1}{%
      doi:\discretionary{}{}{}%
      \nolinkurl{#1}%
    }%
  \endgroup
}
\makeatother

\crefname{equation}{Eq.}{Eqs.}
\crefname{section}{Sec.}{Secs.}
\crefname{table}{Tab.}{Tabs.}
\crefname{figure}{Fig.}{Figs.}
\crefname{subfigure}{Fig.}{Figs.}

\biboptions{sort&compress}

\makeatletter
\def\NAT@spacechar{~}
\makeatother

\graphicspath{{figures/}}

\usepackage{fp}
\usepackage{tikz}
\usetikzlibrary{arrows}
\usetikzlibrary{shapes.geometric}
\usetikzlibrary{patterns}
\usetikzlibrary{decorations.pathreplacing}

\usepackage{threeparttable}

\newcommand*{\partic}{\mathrm{p}}
\newcommand*{\liq}{\mathrm{\ell}}
\newcommand*{\vap}{\mathrm{v}}

\newcommand*{\basefl}{{\mathrm{bf}}}
\newcommand*{\inlet}{{\mathrm{in}}}
\newcommand*{\outlet}{{\mathrm{out}}}

\newcommand{\textcite}[1]{\citet{#1}}
\newcommand{\autocite}[1]{\cite{#1}}

\makeatletter
\newcommand*{\dd}{\@ifnextchar^{\DIfF}{\DIfF^{}}}
\def\DIfF^#1{\mathop{\mathrm{\mathstrut d}}\nolimits^{#1}\gobblesp@ce}
\def\gobblesp@ce{\futurelet\diffarg\opsp@ce}
\def\opsp@ce{%
  \let\DiffSpace\!%
  \ifx\diffarg(%
    \let\DiffSpace\relax
  \else
    \ifx\diffarg[%
      \let\DiffSpace\relax
    \else
      \ifx\diffarg\{%
        \let\DiffSpace\relax
      \fi\fi\fi\DiffSpace}
\makeatother

\newcommand*{\ddx}[1]{\ensuremath{\frac{\dd #1}{\dd x}}}

\newcommand*{\pdx}[1]{\ensuremath{\frac{\partial #1}{\partial x}}}

\renewcommand*{\vec}[1]{\ensuremath{\boldsymbol{#1}}}

\newcommand*{\Nu}{\text{Nu}}

\renewcommand*{\Re}{\text{Re}}
\renewcommand*{\Pr}{\text{Pr}}
\newcommand*{\Gr}{\text{Gr}}

\newcommand*{\Gz}{\text{Gz}}

\usepackage{microtype}
\begin{document}

\begin{frontmatter}

\title{Potential of enhancing a natural convection loop with a thermomagnetically
  pumped ferrofluid}

\author[ER]{Eskil Aursand}
\author[ER]{Magnus Aa. Gjennestad}
\author[ER]{Karl Yngve Lerv{\aa}g\corref{cor}}
\ead{karl.lervag@sintef.no}
\author[ER]{Halvor Lund}

\address[ER]{SINTEF Energy Research, P.O. Box 4671 Sluppen, NO-7465 Trondheim, Norway}
\cortext[cor]{Corresponding author. (Tel.: +4793006297)}

\begin{abstract}
The feasibility of using a thermomagnetically pumped ferrofluid to
enhance the performance of a natural convection
cooling loop is investigated.
First, a simplified analytical estimate for the thermomagnetic pumping
action is derived, and then
design rules for optimal solenoid and ferrofluid are presented.
The design rules are used to set up a
medium-scale (\SI{1}{\meter}, 10-\SI{1000}{\watt}) case study,
which is modeled using a previously published and validated
model~(Aursand et al.~\cite{Aursand15}).
The results show that the thermomagnetic driving force
is significant compared to the
natural convection driving force, and may in some cases greatly surpass it.
The results also indicate that cooling performance can be increased by
factors up to 4 and 2 in the single-phase and two-phase regimes, respectively,
even when taking into the account the added heat from the solenoid.
The performance increases can alternatively be used to obtain a reduction
in heat-sink size by up to \SI{75}{\percent}.
\end{abstract}

\end{frontmatter}

\section{Introduction}
\label{sec:introduction}

Nanofluids are composed of a base fluid, such as water, oil or glycol, with suspended nanoparticles.
Surfactants are often added to improve the stability of the particle suspension and prevent settling and clumping.
Nanofluids have been heavily researched for the last two decades, and a number of potential applications have been proposed \cite{Taylor13}.
In 1995, \textcite{Choi95} showed that nanofluids may have improved conductive and convective heat transfer properties compared to the corresponding base fluid.
That is, nanoparticles increase the thermal conductivity and Nusselt number of the nanofluid.
This increase has been confirmed by a number of authors~\cite{Kakacc09, Xuan00_2, Maiga05}.

If the nanoparticles are magnetizable, the fluid is known as a ferrofluid \cite{Odenbach03}.
Such magnetic nanofluids have a number of interesting applications, one of which is its ability to be thermomagnetically pumped using only a static, inhomogeneous magnetic field and a temperature gradient.
This is sometimes also called magnetocaloric pumping~\cite{Rosensweig85}.
A static magnetic field cannot do any net work by itself.
However, if a thermal gradient is present, a net pumping force can be achieved due to the temperature-dependence of the ferrofluid magnetization.

A cooling device using such a pump would require no moving parts,
something which could provide enhanced reliability, simplicity and
compactness.
Additionally, the thermomagnetic pumping action may increase overall
heat transfer performance in a given geometry, compared to conventional
passive solutions such as natural convection. This may also be used to
obtain more compact solutions with similar performance.

The improved reliability and compactness from replacing a mechanical pump
with a thermomagnetic one may be useful for providing cooling in
remote and hazardous environments, where
deployment and maintenance is expensive and difficult.
Examples include cooling of subsea, space and offshore equipment.
Thermomagnetic pumping is particularly advantageous for space applications, since natural convection cooling systems are not possible in weightless environments.


Since the thermomagnetic pumping force increases with larger temperature
differences, the pump will in some sense be self-regulating.
If the magnetic field is set up by a solenoid or an electro-magnet, the pumping
force may also be externally regulated by adjusting the wire current. Thus a
thermomagnetically pumped cooling system can be externally controlled while also
regulating itself, if necessary.

The concept of using magnetically pumped ferrofluids for heat transfer has been
demonstrated by a number of authors, see e.g.~\textcite{Lian09b}
and \textcite{Xuan11}. In particular, \textcite{Iwamoto11} built an apparatus
for measuring the net driving force of a thermomagnetic pump for different
heat transfer rates and pipe inclinations. \textcite{Yamaguchi13} studied a
magnetically driven system for cooling of microelectromechanical
systems (MEMS). The same group has also studied the effect of magnetic field and
heat flux on the flow rate and pumping power of thermomagnetically pumped
devices~\autocite{Iwamoto12}.
They used a linearized magnetization model and
account for the presence of vapor through a bubble generation
rate.
\textcite{Karimi14} considered a thermomagnetic loop with single-phase
ferrofluid. They used a linearized model for ferrofluid magnetization and
developed a Nusselt-number correlation for the heat transfer. \textcite{Lian09a}
studied a similar system, and they too used a linearized model for ferrofluid
magnetization and analyzed the effects of different factors, such as heat
load, heat sink temperature and magnetic field distribution along the loop on
the heat transfer performance.

While the thermomagnetic pumping effect is confirmed as real, it is
an open question whether the effect is significant and useful in practice
on macroscopic scales
compared to conventional passive solutions such as natural convection.
The purpose of this work is to investigate this with a simulated case-study.

We make use of a recently presented and validated model for
thermomagnetic pumping and heat transfer \cite{Aursand15}. This model treats
phase change in the base fluid and the resulting multiphase flow in a rigorous
manner. It uses thermodynamic equations of state
and includes an advanced ferrofluid magnetization sub-model that is valid
in the whole range from linear Curie regime to the saturation regime.
The model is thus
valid for a large range of parameters, e.g.~a large range of applied external
magnetic field strengths and a large range of ferrofluid temperatures. Having a
general model enables systematic parameter studies of the thermomagnetic heat
transfer performance, as well as optimization of a system design over a large
range of parameters.

In this paper, we suggest a
procedure for designing an optimal solenoid and optimal ferrofluid
for a particular application, and we derive a simple approximation for
the expected thermomagnetic pumping action.
Then we apply the suggested optimization procedures
to set up a case study using the full model from~\cite{Aursand15}.
We consider a \SI{1}{\meter} long natural convection cooling
circuit and show how thermomagnetic pumping can be used to improve its
performance. The natural convection case
will serve as a reference case to which any improvements
obtained by using ferrofluids and/or thermomagnetic pumping will be compared.

In \cite{Aursand15}, it was found that the predicted effect of thermomagnetic pumping was sensitive to heat transfer coefficients, which have high uncertainties.
In the present paper, we eliminate much of this uncertainty by comparing thermomagnetic performance with natural convection, an effect which is also driven by heat transfer.
Any error in heat transfer should then have little effect on the
relative performance enhancements.

In \cref{sec:flow_model}, we briefly present the flow model equations that were introduced in \cite{Aursand15}.
We describe how these equations are solved numerically with periodic boundary conditions for a convection loop.
\Cref{sec:analytical_estimates} derives some analytical estimates of the optimal geometric dimensions of a solenoid and the optimal properties of a ferrofluid for thermomagnetic pumping.
A simplified analytical estimate for the thermomagnetic pumping action is also presented here.
In \cref{sec:case}, we describe the application case that we consider in this paper, with the dimensions of the rig, solenoid, heater and cooler, as well as the properties of the base fluid and particles.
The results from the model with and without thermomagnetic pumping are presented in \cref{sec:results}. These are further discussed in \cref{sec:discussion}.
The focus of the discussion is on how the driving forces vary with temperature, and how nanoparticles and thermomagnetic pumping may improve heat transfer or compactness compared to natural convection of a conventional fluid.
Finally, \cref{sec:conclusions_and_further_work} summarizes to what extent thermomagnetic pumping can improve a natural convection circuit and outlines further work.

\section{Flow model}
\label{sec:flow_model}

In \cite{Aursand15}, we presented a steady-state model for ferrofluid flow in a pipe section. The model includes gravity, magnetic and friction forces and a thermodynamic equation of state.
Here we make use of that model to study a heat transfer loop.

\subsection{Flow model}
The one-dimensional steady-state flow of a ferrofluid may be described by
\begin{gather}
  \ddx{} \left( \alpha_\partic \rho_\partic v \right) = 0,
  \label{eq:flow_steady_pmass}
  \\
  \ddx{} \left( \alpha_\basefl \rho_\basefl v \right) = 0,
  \label{eq:flow_steady_bfmass}
  \\
  \ddx{} \left( p \right) =
  f^{\text{mag}} + f^{\text{fric}} + f^{\text{grav}},
  \label{eq:flow_steady_p}
  \\
  \ddx{}\left( \rho v h \right) =
  v f^{\text{grav}}
  + \dot{q},
  \label{eq:flow_steady_h}
\end{gather}
where $\alpha_k$~(--)
and $\rho_k$~(\si{\kilogram\per\meter\cubed}) are the volume fractions and
densities of the indicated phases, respectively. The subscript $\partic$
describes the particle phase, while the subscript $\basefl$
describes the base fluid phase.
Without a subscript, quantities are total for the phase mixture, such
as the flow velocity $v$~(\si{\meter\per\second}),
mixture density $\rho$~(\si{\kilogram\per\meter\cubed}),
and the mixture specific enthalpy $h$~(\si{\joule\per\kilogram}).

The terms on the right--hand sides of
\eqref{eq:flow_steady_pmass}--\eqref{eq:flow_steady_h} are called the
\textit{source terms}. They are described in detail in~\cite{Aursand15}.
The most novel and important one, the magnetic force term, is given by
\begin{equation}
  f^\text{mag} = \mu_0 M \pdx{H}.
  \label{eq:force_term_mag}
\end{equation}
Here $\mu_0$~(\si{\newton\per\ampere\squared}) is the vacuum permeability,
$H$~(\si{\ampere\per\meter}) is the magnetic field,
and $M$~(\si{\ampere\per\meter}) is the magnetization of the ferrofluid,
\begin{equation}
  M = \chi(H,T) H.
  \label{eq:M_chi_H}
\end{equation}
The factor $\chi$~(--) is the susceptibility of the ferrofluid,
and it generally depends both on the magnetic field and the
temperature $T$~(\si{\kelvin}).

\cref{eq:flow_steady_pmass,eq:flow_steady_bfmass} state the conservation
of particle and base fluid mass, respectively.
\cref{eq:flow_steady_p,eq:flow_steady_h}
describe momentum and energy conservation expressed as ordinary differential
equations (ODEs) in pressure and enthalpy flux.

These equations allow us to integrate along a pipe from an inlet position (subscript ``in'')
to an outlet position (subscript ``out''). In other words, they allow the
mapping of an inlet condition $\vec{u}_\inlet = [p_\inlet, (\rho v h)_\inlet ]$
to an outlet condition $\vec{u}_\outlet$.
The integration is performed along a pipe which may have a variety of components
and orientations. The components and orientations affect the right hand side of the ODEs.
The mass fluxes $\dot{m}_\partic = \alpha_\partic \rho_\partic v$
and
$\dot{m}_\basefl = \alpha_\basefl \rho_\basefl v$
of particles and base fluid are constant along the pipe and need to be
supplied as parameters to the integration procedure.

\subsection{Finding periodic solutions}

In order to describe a loop, we need to add the requirement that the
solution is periodic, i.e.\ $\vec{u}_\inlet = \vec{u}_\outlet$.
While searching for such solutions,
the constants are chosen to be
the pressure and particle volume
fraction at the inlet ($p_\inlet, \alpha_{\partic,\outlet}$),
along with the pipe configuration.

To help find a periodic solution given the above constants,
we define the combined quantity
$\vec{w} \equiv [v_\inlet, h_{\basefl,\inlet}]$, where
$h_\basefl$~(\si{\joule\per\kilogram}) is the base fluid enthalpy.
This $\vec{w}$ is the variable of the search,
with the goal being to find the point which leads
to satisfying $\vec{u}_\inlet = \vec{u}_\outlet$.

We define an objective function
\begin{equation}
  \vec{F}(\vec{w}; p_\mathrm{in}, \alpha_{\partic,\mathrm{in}}) =
  \vec{u}_\inlet - \vec{u}_\outlet,
\end{equation}
which is calculated as follows:
\begin{enumerate}
  \item Based on $\vec{w}$ and $p_\mathrm{in}$
    we perform a thermodynamic pressure-enthalpy equilibrium
    calculation~\cite{Michelsen07} on the base fluid at the inlet.
    This yields the temperature $T$ and base fluid density $\rho_\basefl$.
    Since the density of the particle material $\rho_\partic$ is assumed constant,
    knowing the temperature and pressure also
    allows for the calculation of the particle enthalpy $h_\partic$.
  \item The mass fluxes of particles and base fluid can now be
    calculated at the inlet as
    $\dot{m}_\partic = \alpha_\partic \rho_\partic v$ and
    $\dot{m}_\basefl = (1-\alpha_\partic) \rho_\basefl v$, which will both
    be constant along the pipe.
  \item The enthalpy flux
    $\rho v h = \dot{m}_\partic h_\partic + \dot{m}_\basefl h_\basefl$
    can now be calculated for the inlet.
  \item The values of $\vec{u}_\inlet$,
    $\dot{m}_\partic$ and $\dot{m}_\basefl$ corresponding to
    the current $\vec{w}$ is now known.
    This is the starting point for integrating the flow model
    along the pipe until the outlet.
  \item The flow equations are integrated around the loop
    according to~\cite{Aursand15}. This
    is the dominating contribution to the calculation time of $\vec{F}$.
  \item After the integration, $\vec{u}_\outlet$ is known,
    which allows the calculation of the periodicity errors
    $\vec{u}_\inlet - \vec{u}_\outlet$.
\end{enumerate}

The task is then to solve for the  $\vec{w}$ which gives periodicity,
e.g.\ $\vec{F}(\vec{w}) = 0$. This is a non-linear system of equations which
must be solved numerically.
While the equation system is solved for $\vec{w}$ iteratively,
every single evaluation of $\vec{F}$ involves integrating the ODEs from initial
conditions through a pipe at varying orientations with respect to gravity and with various components such as heaters/coolers
and magnetic field sources.
Thus the Jacobian $\dd{\vec{F}}/\dd{\vec{w}}$ cannot in general be known analytically,
and must be numerically estimated by additional evaluations of $\vec{F}$.

In the present work, we solve $\vec{F}(\vec{w}) = 0$ using \verb|SciPy|'s
\cite{Scipy} wrappers of \verb|MINPACK|’s \cite{More80} \verb|hybrd| and
\verb|hybrj| algorithms for solving non-linear equation systems.

Once a solution for $\vec{w}$ is found, it completely specifies the
inlet/outlet state
when combined with the given $p_\inlet$ and $\alpha_{\partic,\inlet}$.
A final integration of the ODEs from this state then gives the desired properties along the entire loop.

\section{Analysis and estimates}
\label{sec:analytical_estimates}

In this section we provide an analysis of the solenoid geometry, and we present
an estimate for the optimal geometry provided a given set of restrictions.
Further, we derive approximations for the thermomagnetic pump performance in
terms of the pressure increase achieved due to the magnetic force term. This is
used to estimate the optimal particle size distribution for a given ferrofluid.

\subsection{Solenoid}
\label{sub:solenoid}
The solenoid geometry and position is defined by
an inner radius
$R_1$~(\si{\meter}),
an outer radius
$R_2$~(\si{\meter}),
a left end position
$x_1$~(\si{\meter})
and a right end position
$x_2$~(\si{\meter}).
This gives a length $L=x_2-x_1$
and a solenoid thickness $\delta=R_2-R_1$.
For the following analysis, it
is practical to define two aspect ratios:
\begin{align}
  \label{eq:alpha}
  \alpha &\equiv \frac{R_2}{R_1}, \\
  \label{eq:beta}
  \beta &\equiv \frac{L}{2R_1}.
\end{align}
As the solenoid is placed over a given pipe section, it is clear that the inner
radius $R_1$ must be larger than the radius of the contained pipe.

The solenoid consists of a wire of resistivity
$\varrho$~(\si{\ohm\meter}).
The number of turns per length of the solenoid, i.e.\ the turn density, is
$n$~(\si{\per\meter}).
The turns are packed in layers, where for circular wire cross-sections the
optimum packing factor is $\lambda = \pi/(2\sqrt{3}) \approx 0.91$.

\subsubsection{Power consumption}
In order to decide the feasibility of the thermomagnetic pumping with a solenoid
magnet,
the power consumption of the solenoid must the approximated.
The power consumption in a solenoid of constant current is mainly
the ohmic loss in the wire.
If we approximate each wire loop as a circle,
the total wire length $l_\text{wire}$~(\si{\meter}) is
\begin{align}
  l_\text{wire} &= \frac{L n}{\delta} \int_{R_1}^{R_2} 2 \pi r \dd r
  \nonumber \\
    &=\pi n L (2 R_1 + \delta).
  \label{eq:wire_length}
\end{align}
The wire cross-section area $A_\text{wire}$~(\si{\meter\squared}) is
\begin{align}
  A_\text{wire} = \frac{\lambda\delta}{n},
  \label{eq:wire_cross_section}
\end{align}
which combined with \cref{eq:wire_length}
gives a total wire resistance $\mathcal{R}$~(\si{\ohm}) of
\begin{align}
  \mathcal{R} \equiv \frac{\varrho l_\text{wire}}{A_\text{wire}}
  = \frac{\pi \varrho L}{\lambda}
  \left( \frac{\alpha + 1}{\alpha - 1} \right) n^2.
\end{align}
This means that the power consumption
$P$~(\si{\watt}) is
\begin{align}
  P &\equiv \mathcal{R} I^2
    = \varrho \frac{\pi L}{\lambda} \frac{\alpha+1}{\alpha-1}
    (nI)^2.
    \label{eq:power_from_nI}
\end{align}

Note that the wire current density $J$~(\si{\ampere\per\meter\squared}) is
\begin{align}
  J \equiv \frac{I}{A_\text{wire}} = \frac{nI}{\lambda\delta},
  \label{eq:J_from_ni}
\end{align}
which should not be above approximately
\SI{5}{\ampere\per\milli\meter\squared}
in normal wiring.


\subsubsection{Maximum field strength}
\label{ssub:maximum_field_strength}

The magnetic field $H$
along the axis of an empty solenoid of finite length and
width is \cite{Jiles98}
\begin{align}
  \label{eq:H_solenoid_finite_empty}
  H &= \frac{n I}{2(R_2-R_1)}
    \left[(x-x_1) \xi_1 - (x-x_2) \xi_2 \right],
\end{align}
with the spatial derivative
\begin{multline}
  \frac{\dd H}{\dd x}
  = \frac{n I}{2(R_2-R_1)} \Bigg[
    \xi_1 - \xi_2 \\
    + \left(\frac{(x-x_1)^2}{(R_2 + \xi_{21})\xi_{21}}
      - \frac{(x-x_1)^2}{(R_1 + \xi_{11})\xi_{11}}\right) \\
    - \left(\frac{(x-x_2)^2}{(R_2 + \xi_{22})\xi_{22}}
      - \frac{(x-x_2)^2}{(R_1 + \xi_{12})\xi_{12}}\right) \Bigg],
\end{multline}
where
\begin{align}
  \label{eq:H_solenoid_xidef}
  \xi_{ij} &= \sqrt{R_i^2 + (x-x_j)^2},\\
  \xi_j &= \ln\left(\frac{R_2+\xi_{2j}}
                       {R_1+\xi_{1j}}\right).
\end{align}
The $x$-axis is taken to be along the central axis of
the solenoid, directed such that the current runs clockwise when looking in the
positive direction.

The maximum field strength, which occurs at the center of the solenoid, is
\begin{align}
  H_\text{max}
  &= nI \frac{\beta}{\alpha - 1}
  \ln
  \left(
    \frac{\alpha + \sqrt{\alpha^2 + \beta^2} }{1 + \sqrt{1 + \beta^2}}
  \right),
  \label{eq:Hmax_from_ni}
\end{align}
where $\alpha$ and $\beta$ are the aspect ratios defined in \cref{eq:alpha,eq:beta}. $H_\text{max}$ may be given in terms of the power consumption \eqref{eq:power_from_nI},
\begin{align}
  H_\text{max}
  &=
  \sqrt{\frac{\lambda P}{R_1 \varrho}}
  G,
  \label{eq:Hmax_from_Gfactor}
\end{align}
where $G$ is the geometry factor,
\begin{multline}
  G(\alpha, \beta) \equiv
  \sqrt{\frac{\beta}{2\pi (\alpha^2 - 1)}} \\
  \ln
  \left(
    \frac{\alpha + \sqrt{\alpha^2 + \beta^2} }{1 + \sqrt{1 + \beta^2}}
  \right).
\end{multline}
Since $G$ has a global maximum,
\begin{equation}
  G_\text{max} = G(3.0951, 1.8618) = 0.14262,
  \label{eq:G_optimum}
\end{equation}
$H_\text{max}$ reaches its maximum when the inner radius $R_1$ reaches its allowed minimum. For a specified $R_1$, there is a unique optimum for $\alpha$ and $\beta$, and thus for $R_2$ and $L$.

However, one does not always have the luxury of choosing the length of
the solenoid. For instance, in order to achieve maximal temperature change across the
solenoid, it should be a bit longer than the heated part of the pipe.
Thus $L$ is specified by the application. As $R_1$ is already
set as small as possible, $\beta$ is also specified by the application, and so
$G$ will generally not reach its global maximum.
However, for a specified value of $\beta$, $G$ may yet be optimized for the
remaining free geometry parameter $\alpha$. The result will then be the
optimal value of $R_2$ given specified values of $R_1$ and $L$.

\begin{figure}
        \centering
        \includegraphics[width=0.48\textwidth]{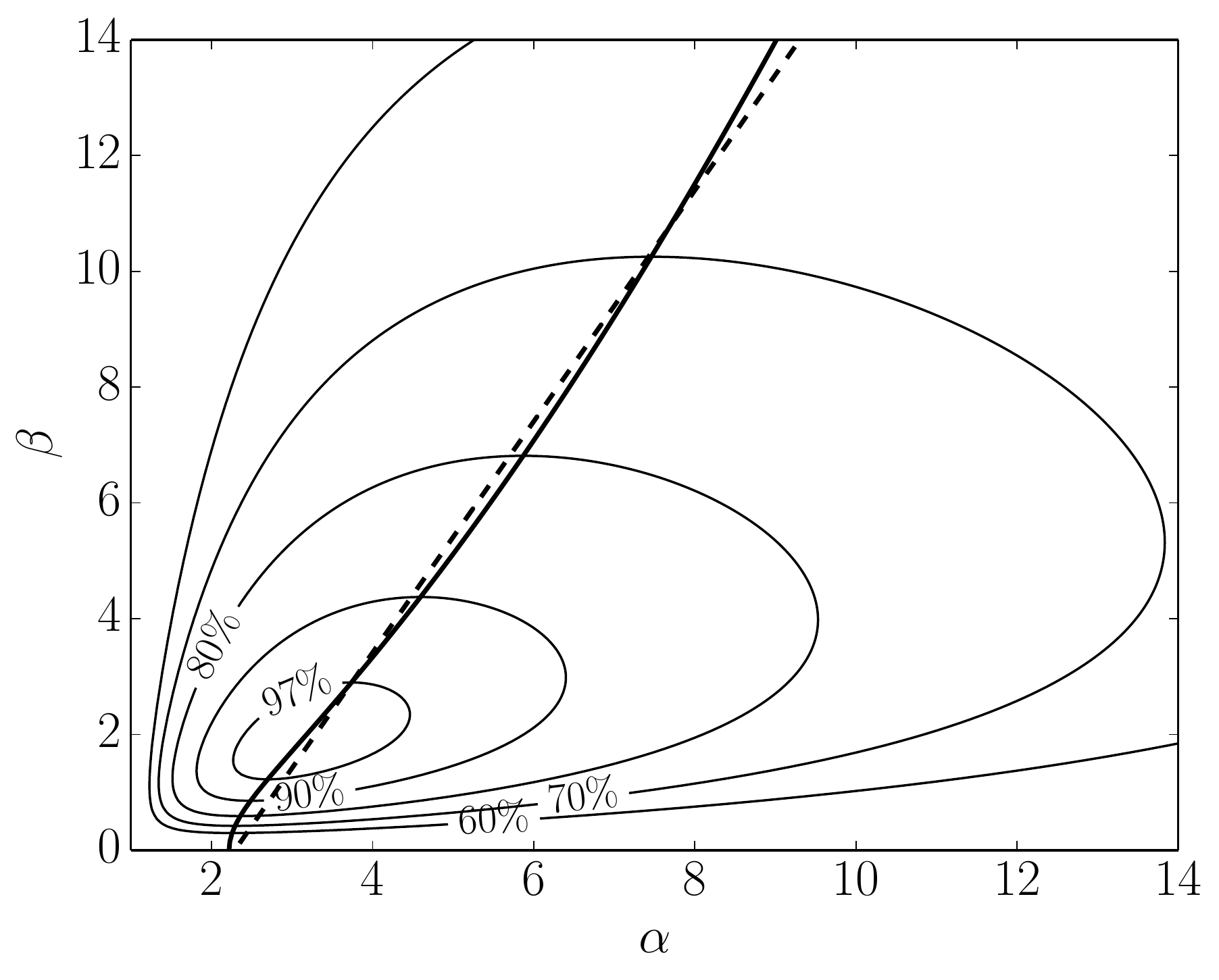}
        \caption{Map of geometry factor $G(\alpha,\beta)$,
          relative to its maximum value. The optimal value of
          $\alpha$ given a value of $\beta$ is shown by the diagonal solid
          line, and a linear fit of that line is shown by the dashed line.}
        \label{fig:map_Gfactor}
\end{figure}

\cref{fig:map_Gfactor} shows the value of $G(\alpha,\beta)$ relative to
its global optimum. It also shows the line of optimal values for a given $\beta$,
which turns out to be approximately linear. The linear fit is
\begin{equation}
  \alpha \approx 0.50\beta + 2.3,
  \label{eq:optim_alpha_given_beta}
\end{equation}
which will be used here as a design rule for obtaining the most field strength
per power consumption, when $L$ and a minimum possible $R_1$ are given.

Another important consideration is the maximum field a solenoid of given
geometry is able to generate before surpassing the maximum allowable wire
current density.
Combining \cref{eq:J_from_ni,eq:Hmax_from_ni} leads to an expression for the
field strength given a wire current density $J$,
\begin{align}
  H_\text{max}
  &=
  J R_1  \beta \lambda
  \ln
  \left(
    \frac{\alpha + \sqrt{\alpha^2 + \beta^2} }{1 + \sqrt{1 + \beta^2}}
  \right).
  \label{eq:Hmax_and_J}
\end{align}

One should note that finding the optimal geometry for maximum field-per-power does not ensure that it is possible to reach a desired field
strength without overstepping the maximum allowable wire current density. A
departure from the optimal geometry may be necessary to reach the desired field
strength with an acceptable current density. In other words, at the cost of
more power per field strength, one may reach any desired field strength.

\subsection{Thermomagnetic pumping}
\label{sub:thermomagnetic_pumping}

The performance of the thermomagnetic pump, in terms of the pressure
increase achieved due to the magnetic force term \eqref{eq:force_term_mag},
may be expressed as
\begin{align}
  \Delta p^\text{mag} &= \int_{-\infty}^{\infty} f^\text{mag} \dd{x}
  \nonumber \\
           &= \mu_0 \int_{-\infty}^{\infty} M \pdx{H} \dd{x}.
  \label{eq:dp_from_MH_area}
\end{align}
In the following, it will be useful to note that the
integral \eqref{eq:dp_from_MH_area} is given by the area inside the
parametric curve $(M(x), H(x))$.

A first approximation of the pump performance may be found by
assuming that most of the temperature change
from the left (upstream) $T_\text{L}$ to
the right (downstream) $T_\text{R}$ happens
within the centre of a solenoid, where the field is approximately
constant and equal to
$H_\text{max}$. In this case, \cref{eq:dp_from_MH_area} reduces to
\begin{multline}
  \Delta p^\text{mag}
  = \mu_0 \int_{0}^{H_\text{max}}
    \big[M(H,T_\text{L}) \\ - M(H,T_\text{R})\big]\dd{H},
  \label{eq:dp_from_MH_area_center_heating}
\end{multline}
which is simply the area between the magnetization curves $M(H)$ at $T_\text{L}$ and $T_\text{R}$ up to $H_\text{max}$. This fact will be used in the discussion of the next section.

A second approximation may be found in the linear magnetization regime, where the magnetic field is weak and/or the temperature is high. Here one may assume that the susceptibility only depends on the temperature, that is, $\chi\equiv\chi(T)$. Thus, from \cref{eq:dp_from_MH_area_center_heating,eq:M_chi_H}, we get
\begin{align}
  \Delta p^\text{mag}
  &= \mu_0 \int_{0}^{H_\text{max}}
  \left[\chi(T_\text{L}) - \chi(T_\text{R}) \right] H \dd{H}
  \nonumber \\
  &= \frac{\mu_0 }{2} \Delta \chi H_\text{max}^2,
  \label{eq:dp_approx_from_delta_chi}
\end{align}
where $\Delta \chi \equiv \chi(T_\text{L}) - \chi(T_\text{R})$,
i.e.\ the change in magnetic susceptibility across the solenoid
from left (upstream) to right (downstream).
Again we remark that the integral is given by the area inside the $H$--$M$ curve, in this case
as a triangle with vertices $(0,0)$, $(H_\text{max}, M_\text{L})$
and $(H_\text{max}, M_\text{R})$.

If we combine \cref{eq:Hmax_from_Gfactor,eq:dp_approx_from_delta_chi}, we can express the pumping performance only in terms of power consumption, solenoid properties, and change in ferrofluid susceptibility,
\begin{equation}
  \Delta p^\text{mag} = \frac{\mu_0 }{2} \Delta \chi
    \frac{\lambda G^2}{R_1 \varrho} P.
    \label{eq:dp_from_P}
\end{equation}

%
%
%

\subsection{Ferrofluid design}
\label{sub:ferrofluid_design}

In the following, we will estimate a particle size distribution that gives an
optimal thermomagnetic pumping effect for a given particle material, magnetic
field and temperature range. We also provide some general design remarks and
give an estimate for the pumping effect for a given case.

First, we note that the susceptibility of a ferrofluid with a particle volume
fraction $\alpha_\partic$ is given by
\begin{equation}
  \chi = \alpha_\partic \chi_\partic =
  \alpha_\partic \frac{M_\partic(H,T)}{H},
  \label{eq:chi_from_Mp}
\end{equation}
where $M_\partic$ is the average particle magnetization.

In~\cite{Aursand15},
we present two different models for $M_\partic$: The first assumes a uniform
particle distribution (monodisperse), whereas the second uses a Gaussian
particle distribution.  The monodisperse model is used here, as it
requires less computational work at the expense of accuracy. The model
may be summarized as
\begin{multline}
  M_\partic(V_\partic,H,T) \\
  = M_\text{sat} (T)
    \mathcal{L}\left(\frac{\mu_0 V_\partic M_\text{sat}(T) H}{k_B T} \right),
  \label{eq:mag_p_langevin}
\end{multline}
where $V_\partic$ (\si{\meter\cubed}) is the volume of a single particle,
$\mathcal{L}$~is the \emph{Langevin function}, and the
saturation magnetization $M_\text{sat} (T)$
is modeled linearly to reach zero beyond a given Curie temperature,
see~\cite{Aursand15}.
Note that the Langevin function is approximately linear for small values of its argument.
In this limit, the monodisperse model \eqref{eq:mag_p_langevin} reduces to
what is called the \textit{Curie regime} or \textit{Curie law}.

Given a particle material, represented by the function $M_\text{sat}(T)$,
a maximum field $H_\text{max}$
and a temperature range $T_\text{L}$ to $T_\text{R}$,
we may use the particle
magnetization model \eqref{eq:mag_p_langevin} to optimize the
thermomagnetic pumping performance $\Delta p^\text{mag}$ as a function of particle size through
\cref{eq:dp_from_MH_area_center_heating}.
Here the particle concentration is only a factor between $M_\partic$ and $M$,
and so it does not affect the optimal particle size.
It may thus be adjusted as desired
after optimization.

From the optimization through \cref{eq:dp_from_MH_area_center_heating}, we find that the optimal
particle size decreases with increasing $H_\text{max}$ and increases with
increasing $T_\text{R}-T_\text{L}$. Further, we find that
the magnetisation curves given an optimal particle size are
such that $H_\text{max}$ is well beyond the Curie regime, with a magnetisation
about \SI{90}{\percent} of saturation. This means that, for $T_\text L < T < T_\text R$,
\begin{equation}
  M_\partic(V_p^\text{optim},H_\text{max},T) \simeq 0.9 M_\text{sat} (T).
\end{equation}

\begin{figure}
        \centering
        \includegraphics[width=0.45\textwidth]{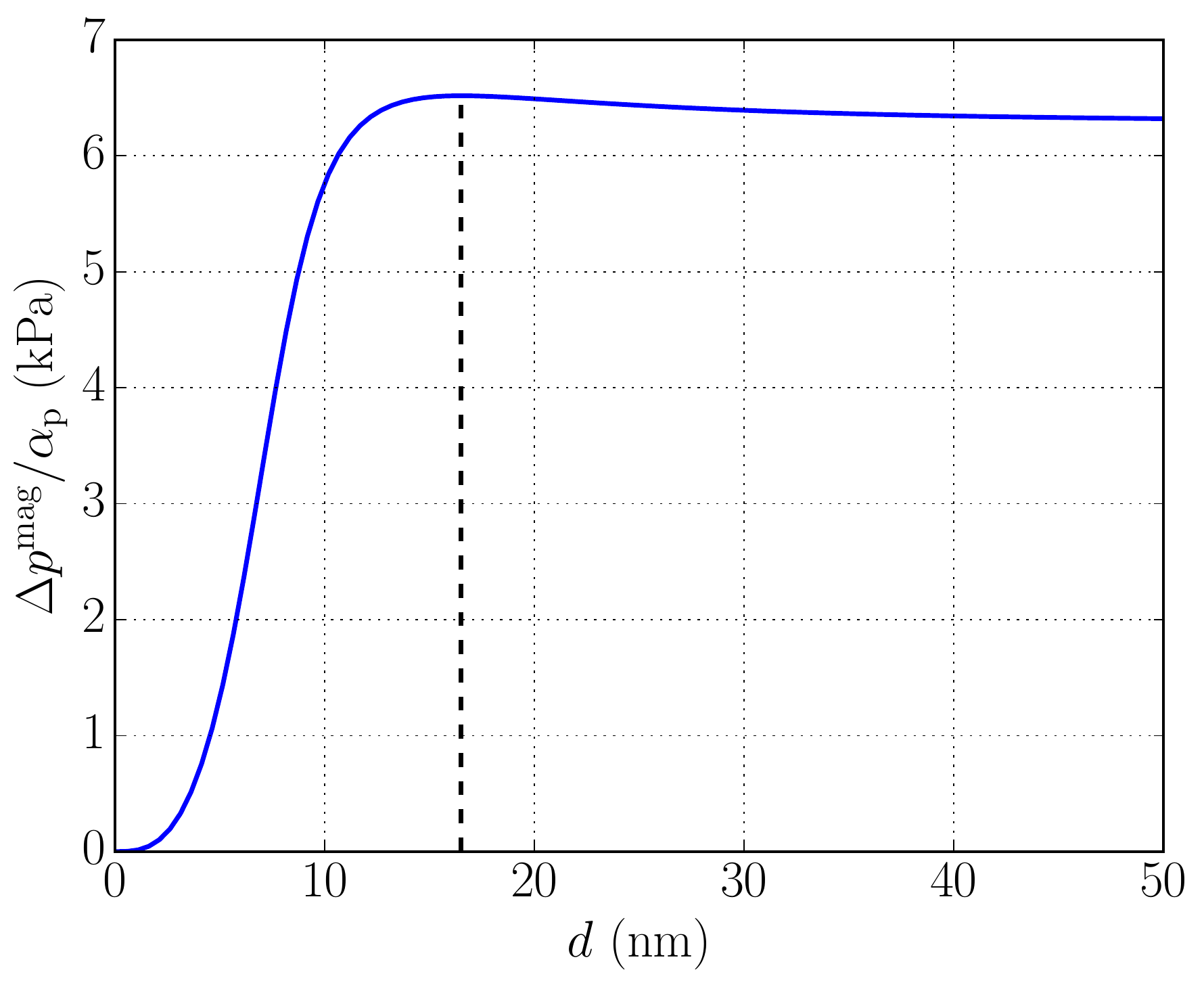}
        \caption{The estimated thermomagnetic pumping performance as a
        function of particle size, as given by
        \cref{eq:dp_from_MH_area_center_heating}.
        The particle volume fraction $\alpha_\partic$ does not affect
        the optimization, and one may thus multiply with any chosed value
        to obtain the final $\Delta p^\text{mag}$.
        The particle material
        is the same as the one chosen in the main case study,
        and the optimization was performed for
        $H_\text{max} = \SI{100}{\kilo\ampere\per\meter}$,
        $T_\text{L} = \SI{10}{\celsius}$ and
        $T_\text{R} = \SI{50}{\celsius}$. The dashed line marks the
        identified optimum.
        }
        \label{fig:d_opt}
\end{figure}

An example of such an optimization may be seen in \cref{fig:d_opt}.
Notice how $\Delta p^\text{mag}$ varies little for particle sizes that
are close to the optimum. This means that there is considerable leeway for
deviating from the optimum, for practical reasons such as stability.
In particular, we find that increasing the particle size will not decrease performance much. However,
decreasing the particle size will eventually lead to a rapid decrease in
$\Delta p^\text{mag}$.

Provided that this optimization is performed,
obtaining a large $\Delta p^\text{mag}$ then depends on:
\begin{itemize}
  \item Having particles with a large saturation magnetization and a low
    Curie temperature.
  \item Having a large $\alpha_\partic$, which may be limited by stability
    or viscosity concerns.
\end{itemize}
Aside from ferrofluid properties,
the mechanisms giving efficient thermomagnetic pumping are then
\begin{itemize}
  \item Increasing temperature from left to right, which will decrease both
    $1/T$ and $M_\text{sat}$ from left to right.
  \item Decreasing $\alpha_\partic$ from left to right. This may be achieved
    by a decrease in the base fluid density, which through the conservation of
    base fluid mass flow will cause an increase in the velocity. An increase in
    the velocity will cause a decrease in $\alpha_\partic$, through the
    conservation of particle mass flow. In single-phase flow, this may be caused
    by the slight thermal expansion of the liquid, but this effect
    mainly becomes significant when boiling occurs, which will drastically reduce
    the base fluid density.
\end{itemize}

Finally, it is of interest to estimate the magnitude of the thermomagnetic
pumping force under optimal conditions. We use the approximation from
\cref{eq:dp_approx_from_delta_chi}. The largest possible value is obtained when
the susceptibility $\chi$ on the downstream side of the solenoid vanishes, in
which case $\Delta\chi$ is simply equal to $\chi$ on the upstream side. As
noted above, the magnetization is about \SI{90}{\percent} of saturation at $H_\text{max}$.
Thus, given some realistic values, $H_\text{max}
= \SI{100}{\kilo\ampere\per\meter}$, $M_\text{sat}(T_\text L)
= \SI{200}{\kilo\ampere\per\meter}$ and $\alpha_\partic=0.1$, we get an
estimate for the pumping force, $\Delta p^\text{mag} \sim
\SI{1}{\kilo\pascal}$.


\section{Case specification}
\label{sec:case}

This section describes the setup of the case-study. The case is chosen such as
to demonstrate the feasibility of enhancing a simple natural-convection based
passive cooling concept by adding magnetic nanoparticles and a static magnetic
field.

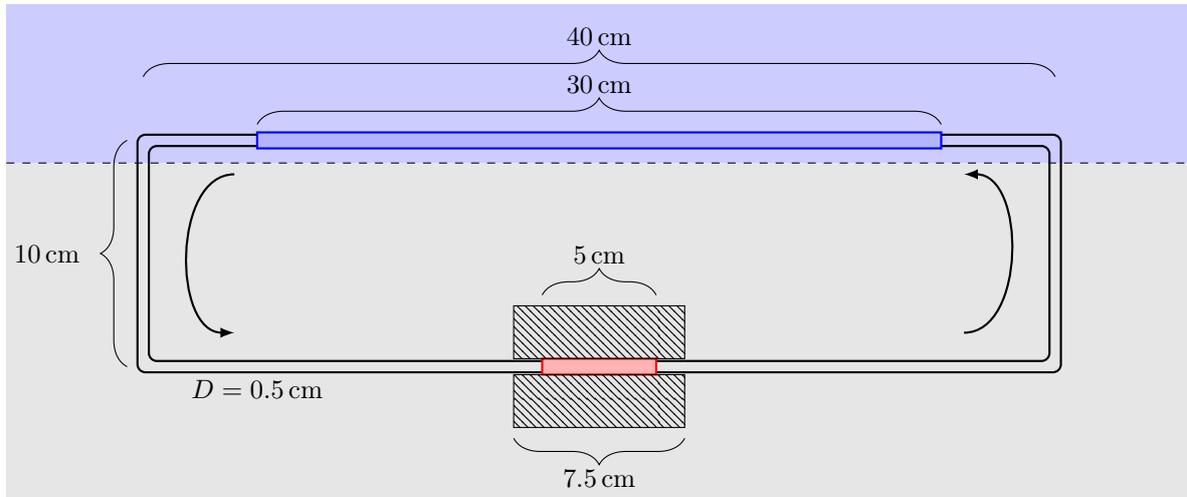
\begin{figure*}[t]
  \centering
  \begin{tikzpicture}
    [
      scale=0.3
    ]

    \def\width{40}
    \def\height{10}
    \def\diam{0.5}
    \def\heaterlength{5}
    \def\coolbundlelength{30}
    \FPeval\bundlediam{1.414*\diam} 

    \def\solL{7.5}
    \FPeval\solRin{0.5*\bundlediam}
    \FPeval\solDin{2.0*\solRin}
    \FPeval\beta{\solL/\solDin}
    \FPeval\alpha{0.50*\beta + 2.3}
    \FPeval\solRout{\alpha*\solRin}

    \def\avar{5.0}
    \FPeval\res{2.0*\avar}

    \def\figborder{6}

    \tikzstyle{pipewall} = [thick, rounded corners=6*\diam]

    \fill[gray!20] (-\figborder, -\figborder) rectangle
      (\width+\figborder, \height-2*\diam);
    \fill[blue!20] (-\figborder, \height-2*\diam) rectangle
      (\width+\figborder, \height+\figborder);
    \draw[dashed] (-\figborder, \height-2*\diam)
      -- (\width+\figborder, \height-2*\diam);

    \draw[pipewall] (0.5*\width-0.5*\coolbundlelength,\height+0.5*\diam)
      -- (-0.5*\diam,\height+0.5*\diam)
      -- (-0.5*\diam,-0.5*\diam) -- (0.5*\width-0.5*\heaterlength,-0.5*\diam);
    \draw[pipewall] (0.5*\width-0.5*\coolbundlelength,\height-0.5*\diam)
      -- (0.5*\diam,\height-0.5*\diam)
      -- (0.5*\diam,+0.5*\diam) -- (0.5*\width-0.5*\heaterlength,+0.5*\diam);
    \draw[pipewall] (\width-0.5*\width+0.5*\coolbundlelength,\height+0.5*\diam)
      -- (\width+0.5*\diam,\height+0.5*\diam)
      -- (\width+0.5*\diam,-0.5*\diam)
      -- (\width-0.5*\width+0.5*\heaterlength,-0.5*\diam);
    \draw[pipewall] (\width-0.5*\width+0.5*\coolbundlelength,\height-0.5*\diam)
      -- (\width-0.5*\diam,\height-0.5*\diam)
      -- (\width-0.5*\diam,0.5*\diam)
      -- (\width-0.5*\width+0.5*\heaterlength,0.5*\diam);

    \draw[thick, color=blue, fill=blue!30]
      (0.5*\width-0.5*\coolbundlelength,\height-0.5*\bundlediam)
         rectangle (\width - 0.5*\width + 0.5*\coolbundlelength,
                    \height+0.5*\bundlediam);
    \draw[thick, color=red, fill=red!30]
      (0.5*\width-0.5*\heaterlength,-0.5*\bundlediam)
         rectangle (\width - 0.5*\width + 0.5*\heaterlength,
                     0.5*\bundlediam);

    \draw[pattern=north west lines] (0.5*\width-0.5*\solL, \solRin) rectangle
      (0.5*\width+0.5*\solL, \solRout);
    \draw[pattern=north west lines] (0.5*\width-0.5*\solL, -\solRout) rectangle
      (0.5*\width+0.5*\solL, -\solRin);



   \node[] at (5.0,-1.0)  {$D=\SI{\diam}{cm}$};

   \draw [decorate,decoration={brace,amplitude=10pt},xshift=-20pt,yshift=0pt]
   (0.0,0.0) -- (0.0,\height)
   node [black,midway,xshift=-30pt] {\SI{\height}{\centi\meter}};

   \draw [decorate,decoration={brace,amplitude=10pt},xshift=0pt,yshift=80pt]
   (0.0,\height) -- (\width,\height)
   node [black,midway,yshift=15pt] {\SI{\width}{\centi\meter}};

   \draw [decorate,decoration={brace,amplitude=10pt},xshift=0pt,yshift=20pt]
   (0.5*\width-0.5*\coolbundlelength,\height) --
   (\width - 0.5*\width + 0.5*\coolbundlelength,\height)
   node [black,midway,yshift=15pt]
   {\SI{\coolbundlelength}{\centi\meter}};

   \draw [decorate,decoration={brace,amplitude=10pt},xshift=0pt,yshift=90pt]
   (0.5*\width-0.5*\heaterlength,0.0) --
   (\width - 0.5*\width + 0.5*\heaterlength,0.0)
   node [black,midway,yshift=15pt] {\SI{\heaterlength}{\centi\meter}};

   \draw [decorate,decoration={brace,amplitude=10pt},xshift=0pt,yshift=-90pt]
   (0.5*\width+0.5*\solL,0.0) --
   (0.5*\width-0.5*\solL,0.0)
   node [black,midway,yshift=-15pt] {\SI{\solL}{\centi\meter}};

  \draw[thick,-latex] (0.9*\width, 0.15*\height) to[out=0, in=0] (0.9*\width,0.85*\height);
  \draw[thick,-latex] (0.1*\width, 0.85*\height) to[out=180, in=180] (0.1*\width,0.15*\height);

  \end{tikzpicture}
  \caption{A schematic of the simulated flow loop, showing the heat source at the
    bottom part and the heat sink at the top part. The solenoid cross-section
    is shown as hatched boxes. Gravity points downward in-plane.
    All dimensions are to scale. Arrows show chosen flow direction.}
  \label{fig:benchmark_rig}
\end{figure*}

The case consists of a flow loop that connects a heat source to a heat sink,
where a solenoid is placed over the heat source, see \cref{fig:benchmark_rig}.
At the lower left corner, we specify an initial pressure $p_\inlet
= \SI{1.0}{atm}$. The case may be thought of as the cooling of a hot device
inside a larger object, with some cold reservoir, e.g. ocean water, available
on the outside. We call these elements the \emph{heater} and the \emph{cooler},
respectively.

Note that the case is symmetric, and so will the natural convection and
magnetic forces be if the fluid state is uniform around the loop.  However,
this equilibrium is unstable, and any beginning flow in one direction will
induce stronger flow in the same direction. Exactly how to initiate this
symmetry break is not the focus of this study, thus a counterclockwise flow is
assumed.

In the following, we describe in detail how the different parts of the case are
modeled and defined. First, in \cref{sub:case_flowloop} we describe the flow
loop and the friction models. Next, in \cref{sub:case_htc}, we describe how the
heater and cooler are modeled. In \cref{sub:case_solenoid}, we explain how the
solenoid is designed by use of the previously discussed design rules. And
finally, in \cref{sub:case_ferrofluid} we describe the ferrofluid. In
particular, we use the earlier results to get optimal nanoparticle sizes for
the current case.

\subsection{Flow loop}
\label{sub:case_flowloop}

The flow loop is modeled as a simple pipe with diameter $D=\SI{0.5}{cm}$, except
at the heater and cooler where it is modeled as a bundle of ten smaller pipes of
the same total flow cross section. The intruduction of tube bundles increases
the fluid-to-wall heat transfer area, resulting in more heat transferred to the
fluid and larger thermomagnetic and natural convection driving forces, at the
expense of increased friction.  The bundles are set to cover a cross section two
times as large as the flow cross section, to make room for any (not currently
modeled) heat exchanger cross flow. This increases the total pipe radius at the
bundles by a factor of $\sqrt{2}$, which becomes the restriction on the inner
radius of the solenoid.

The flow throughout the loop is mostly laminar, although some regions may have
turbulent flow. We therefore use a friction factor model that depends on the
local Reynolds number.  For low Reynolds numbers, below 2100, we use a simple
laminar model as described in \cite{Aursand15}. For large Reynolds numbers,
above 4000, we use the correlation by \textcite{Haaland83}. In the transition
regime, that is for Reynolds numbers in the range 2100--4000, we use a smooth
interpolation between the laminar and the turbulent friction models.

We remark that the Reynolds number will be approximately a constant multiplied
with the local tube diameter along the flow loop, thus lower inside the
heater/cooler than through the main loop. This follows because the total flow
cross-section is constant throughout the rig, and so the mass flux is constant.
Further, the ferrofluid viscosity is mostly constant, since the base fluid
viscosity is constant.

\subsection{Heater and cooler}
\label{sub:case_htc}

For simplicity, the inner wall temperatures at the heater and cooler are kept
constant. Herein lies an implicit assumption of no thermal resistance between
the walls and external constant-temperature reservoirs.

The local heat transfer coefficients (HTCs) are calculated from the difference
between the wall temperature and the local average fluid temperature. The
particular HTC model or correlation that we use depends locally on the fluid,
its state, and the temperature difference.

If there is no boiling or condensation, either at the heater or the cooler, the
choice of HTC model depends on whether there are particles in the fluid or
not. In the former case, the laminar Nusselt number correlation developed by
\textcite{Xuan03}, shown in \cite[Eq.~6]{Kakacc09}, is used. In the latter case,
the perfectly laminar Nusselt number of 3.66 is used.
If there is boiling in the heater, sub-cooled or
saturated, the correlation of \textcite{Chen66} is used. Sub-cooled boiling is
initiated once the heater temperature is above the bubble temperature at the
local pressure.  For condensation at the cooler, the correlation by
\textcite{Boyko67} is used. For boiling and condensation, the same correlations
are used whether there are particles present or not. However, in the latter
case, the ferroliquid properties are used instead of the pure liquid properties
as input to the correlations.

\subsection{Solenoid}
\label{sub:case_solenoid}
In order to achieve an optimal thermomagnetic pumping force, we need as high a
temperature difference between the peaks of $\dd{H}/\dd{x}$ as possible.
Due to the fact that the thermomagnetic pump drives flow from a cold inlet
to a hot outlet, it will only work if the field is at the heater, not the
cooler. The solenoid must thus be placed around the heater. We use a solenoid
length of \SI{7.5}{\centi\meter} so that it fully covers the heater.

As explained in \cref{sub:solenoid}, given a length, the optimal field per
power is
achieved when the inner radius is as small as possible. This means wrapping it
around the heater bundle of a diameter which is $\sqrt{2}$ times
the main pipe diameter, i.e.\ \SI{0.71}{\centi\meter}.
Using \cref{eq:optim_alpha_given_beta}, the outer diameter should then be
\SI{5.38}{\centi\meter}. This corresponds to $\alpha=7.60$ and
$\beta=10.6$, which as seen in \cref{fig:map_Gfactor} gives about
\SI{70}{\percent} efficiency with respect to the field strength that would have been obtained
if the the solenoid length were a free parameter.
The resulting solenoid geometry is illustrated to scale
in \cref{fig:benchmark_rig}.

According to \cref{eq:Hmax_and_J}, this solenoid may reach fields of
approximately \SI{100}{\kilo\ampere\per\meter} before surpassing the
allowed wire current density, and therefore the cases will be run with fields
only up to this value.

\subsection{Ferrofluid}
\label{sub:case_ferrofluid}
The ferrofluid that we consider here is a mixture of a commercial ferrofluid
(``TS50K'') and $n$-Hexane. This is similar to the ferrofluid that was used in
the experiments by~\citet{Iwamoto11}, as well as in the validation of our
model~\cite{Aursand15}.

In \cref{ssub:base_fluid,ssub:particles}, respectively, we describe in detail
how the base fluid and particle properties are set. The properties are listed
and summarized in~\cref{tab:fluid_properties}.

\begin{table}[htbp]
  \centering
  \caption{List of ferrofluid properties divided into model parameters and
    derived properties. (*): Standard ambient temperature and pressure
    (\SI{300}{\kelvin}, \SI{1}{atm}), where no vapor is present.}
  \label{tab:fluid_properties}
  \begin{tabular}{lll}
    \toprule
    Description     &   & Value    \\
    \midrule
    \textbf{Model parameters}   &    &  \\
    \textit{Base fluid} & & \\
    Mol.\ frac.\ decane & $z_1$  & 0.5 \\
    Mol.\ frac.\ hexane & $z_2$  & 0.5 \\
    Liquid viscosity & $\eta_\liq$ & \SI{5.31e-4}{\pascal\second}  \\
    Vapor viscosity  & $\eta_\vap$ & \SI{7.65e-6}{\pascal\second} \\
    Liquid conductivity  & $\lambda_\liq$
                             & \SI{0.130}{\watt\per\meter\per\kelvin} \\
    Vapor conductivity   & $\lambda_\vap$
                             & \SI{0.0202}{\watt\per\meter\per\kelvin} \\
    Surface tension  & $\sigma_\basefl$
                              & \SI{0.0120}{\newton\per\meter}\\
    \textit{Nanoparticles} & & \\
    Density & $\rho_\partic$  & \SI{5000}{\kilogram\per\meter\cubed} \\
    Specific heat capacity &  $c_\partic^p$
                           & \SI{750}{\joule\per\kilogram\per\kelvin} \\
    Sat. mag.\ at \SI{300}{\celsius} & $M_\text{sat}^\circ$
                            & \SI{250.0}{\kilo\ampere\per\meter} \\
    Curie temperature  & $T_\text{C}$ & \SI{500}{\kelvin} \\
    Conductivity & $\lambda_\partic$  &  \SI{29.0}{\watt\per\meter\per\kelvin} \\
    Particle diameter & $d_\partic$  & \SI{10.0}{\nano\meter} \\
    Inlet volume fraction & $\alpha_{\partic,\inlet}$ & \SI{10}{\percent}  \\
    \textbf{Derived properties}   & & \\
    \textit{Base fluid} & & \\
    Density (*)  & & \SI{700.2}{\kilogram\per\meter\cubed}\\
    Heat. cap. (*) & & \SI{2252}{\joule\per\kilogram\per\kelvin} \\
    Vol. heat. cap. (*) & & \SI{1576}{\kilo\joule\per\meter\cubed\per\kelvin} \\
    Bubble point (\SI{1}{atm}) & & \SI{91.6}{\celsius}\\
    Latent heat (\SI{1}{atm}) & & \SI{320.4}{\kilo\joule\per\kilogram} \\
    \textit{Ferroliquid} ($\alpha_\partic=0.1$) & & \\
    Density (*) & & \SI{1130}{\kilogram\per\meter\cubed}\\
    Heat. cap. (*) & & \SI{1626}{\joule\per\kilogram\per\kelvin} \\
    Vol. heat. cap. (*) & & \SI{1837}{\kilo\joule\per\meter\cubed\per\kelvin} \\
    Viscosity  & & \SI{6.91e-4}{\pascal\second}\\
    Conductivity  & & \SI{0.173}{\watt\per\meter\per\kelvin} \\
    \bottomrule
  \end{tabular}
\end{table}

\subsubsection{Base fluid}
\label{ssub:base_fluid}
The base fluid of TS50K is kerosene, which is a mixture of more than 20
components. As in~\cite{Aursand15}, we use undecane to represent the kerosene
so that the base fluid of the ferrofluid is a binary hydrocarbon mixture of
hexane and decane. The hexane and decane were mixed in equal molar amounts,
which corresponds to \SI{62}{\percent} decane and \SI{38}{\percent} hexane on a mass basis. We note that
the hexane is more volatile than the decane.

For a given temperature and pressure, the thermodynamic properties of the
mixture, such as density, heat capacity, enthalpy, latent heat and
liquid--vapor equilibrium, are predicted from the thermodynamic equation of
state based on these components and their relative amounts.
This process is described in~\cite{Aursand15}.

We used data from \cite{NISTwebbook} for the transport properties of the
components. The liquid properties were taken at \SI{1}{atm} and
\SI{20}{\celsius}, while the vapor properties were taken at the component
boiling point at \SI{1}{atm}. The transport properties of the components were
combined into base fluid mixture properties for each phase based on the
overall composition fractions. For the mixture viscosity, we used the Grunberg
and Nissan equation~\cite{Grunberg49}, and for the mixture conductivity, we
used the Filippov equation~\cite{Filippov68}. The mixture surface tension was
calculated as a simple molar fraction weighted average of the two components.

The viscosity and conductivity of the ferroliquid phase (liquid phase and particles) is
a modification of the base liquid properties according to correlations
stated in~\cite{Aursand15}. The surface tension is assumed unmodified by the
presence of nanoparticles.

\subsubsection{Particles}
\label{ssub:particles}
The ferrofluid is made of \ce{MnZn}-Ferrite particles.
Exact values for the density and heat capacity are difficult to
know precisely, as they will depend on the relative amounts of \ce{Mn} and \ce{Zn}.
The values
will also be different in nanoparticle form compared to bulk values.
Reasonable values close to the ones used in~\cite{Aursand15} were chosen.
The value for thermal conductivity was taken from~\cite{Lee00}.

The material magnetization properties $M_\text{sat}^\circ$ and $T_\text{C}$
were set in the reasonable region found when fitting to experimental
\ce{MnZn}-Ferrite ferrofluid magnetization data~\cite{Aursand15}.

\begin{figure}
        \centering
        \includegraphics[width=0.48\textwidth]{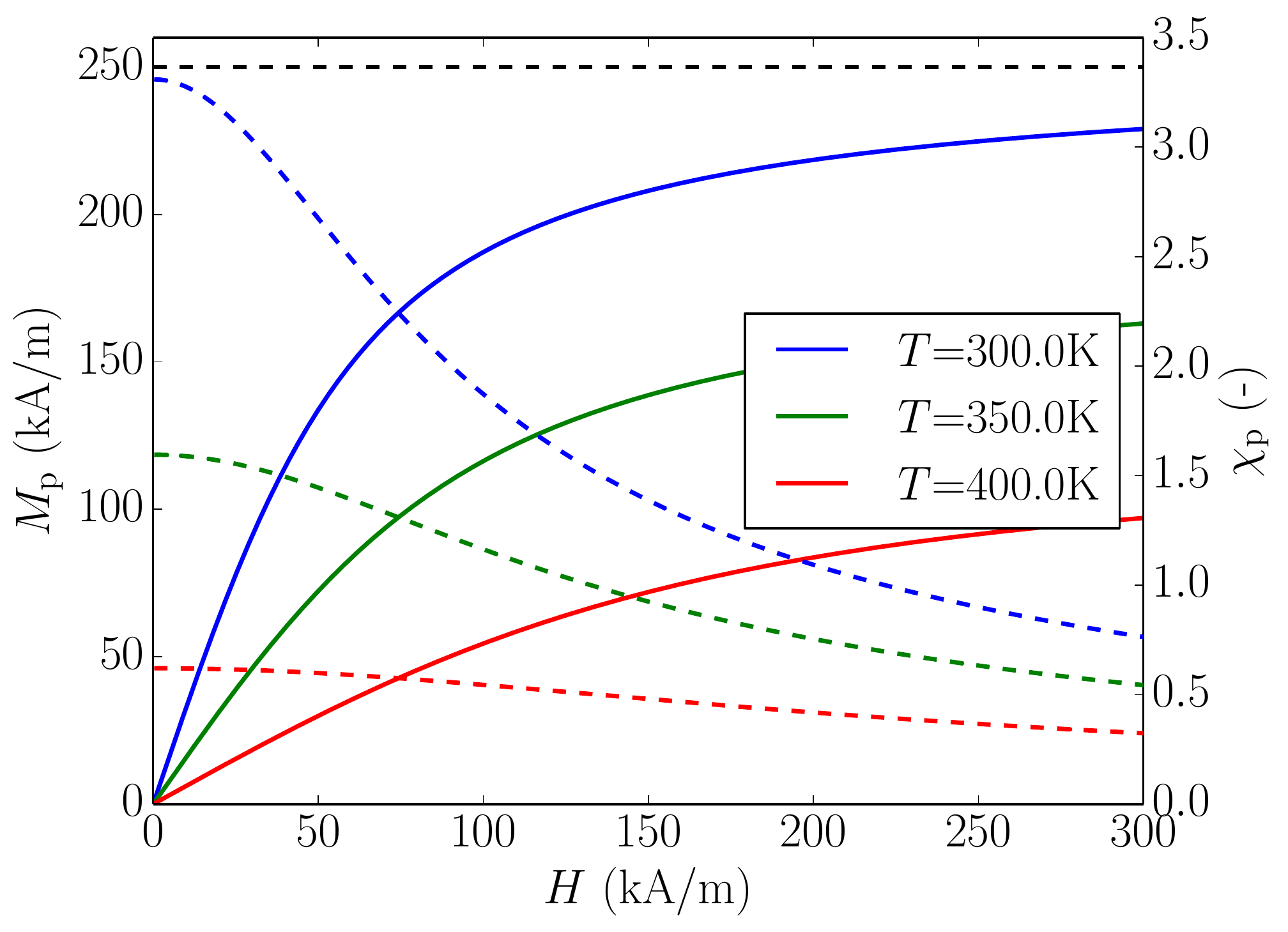}
        \caption{The magnetization curves (solid lines, left axis) and
          susceptibilities (dashed lines, right axis)
                 of the particle ensemble, given
                 the parameters in~\cref{tab:fluid_properties}, at
                 different temperatures. The horizontal dashed line marks the saturation
                 magnetization at \SI{300}{\kelvin}.
                 }
        \label{fig:mag_curves}
\end{figure}

The remaining particle property is the diameter.
An optimization as described in \cref{sub:ferrofluid_design} was performed,
and an optimum was found at about \SI{16}{\nano\meter}, as seen in
\cref{fig:d_opt}.
However, all sizes
above \SI{10}{\nano\meter} showed $\Delta p^\text{mag}$ values above
about \SI{90}{\percent} of the optimum,
and were thus considered good candidates. To avoid stability concerns, and
because it is a common particle size for ferrofluids, a size of
\SI{10}{\nano\meter} was chosen for this study.
The magnetization curves resulting from these parameters may be seen
in \cref{fig:mag_curves}.

The achievable $\Delta \chi$, and thus also the potential performance of
the thermomagnetic pump, is expected to scale with the particle volume
fraction. However, to ensure the practical relevance of the results obtained,
the inlet volume fraction $\alpha_{\partic,\inlet}$ is kept at a moderate \SI{10}{\percent},
which is in the range of common commercial ferrofluids.

\section{Results}
\label{sec:results}

We use the procedure and model equations described in \cref{sec:flow_model} to
obtain steady periodic solutions for the case described in the previous
section. The results from the simulations are presented here and will be
further discussed in the next section.

Two variables were varied:
\begin{itemize}
  \item $\Delta T$: The temperature difference between the heater and cooler,
    which is changed by changing the heater temperature. The cooler was kept at
    $T_\text{cold} = \SI{10}{\celsius}$, to simulate heat exchange against
    seawater.
  \item $H_\text{max}$: The maximum field strength provided by the solenoid,
    which relates to the solenoid power dissipation according to
    \cref{eq:Hmax_from_Gfactor}.
\end{itemize}

Since the flow loop is vertical, a driving force from natural convection
will always be present. This effect alone serves as a reference for the
following results. The reference case was run across the range of $\Delta T$,
with no particles and no field, giving conventional natural convection with
a conventional fluid.

\begin{figure}[tb]
        \centering
        \includegraphics[width=0.48\textwidth]{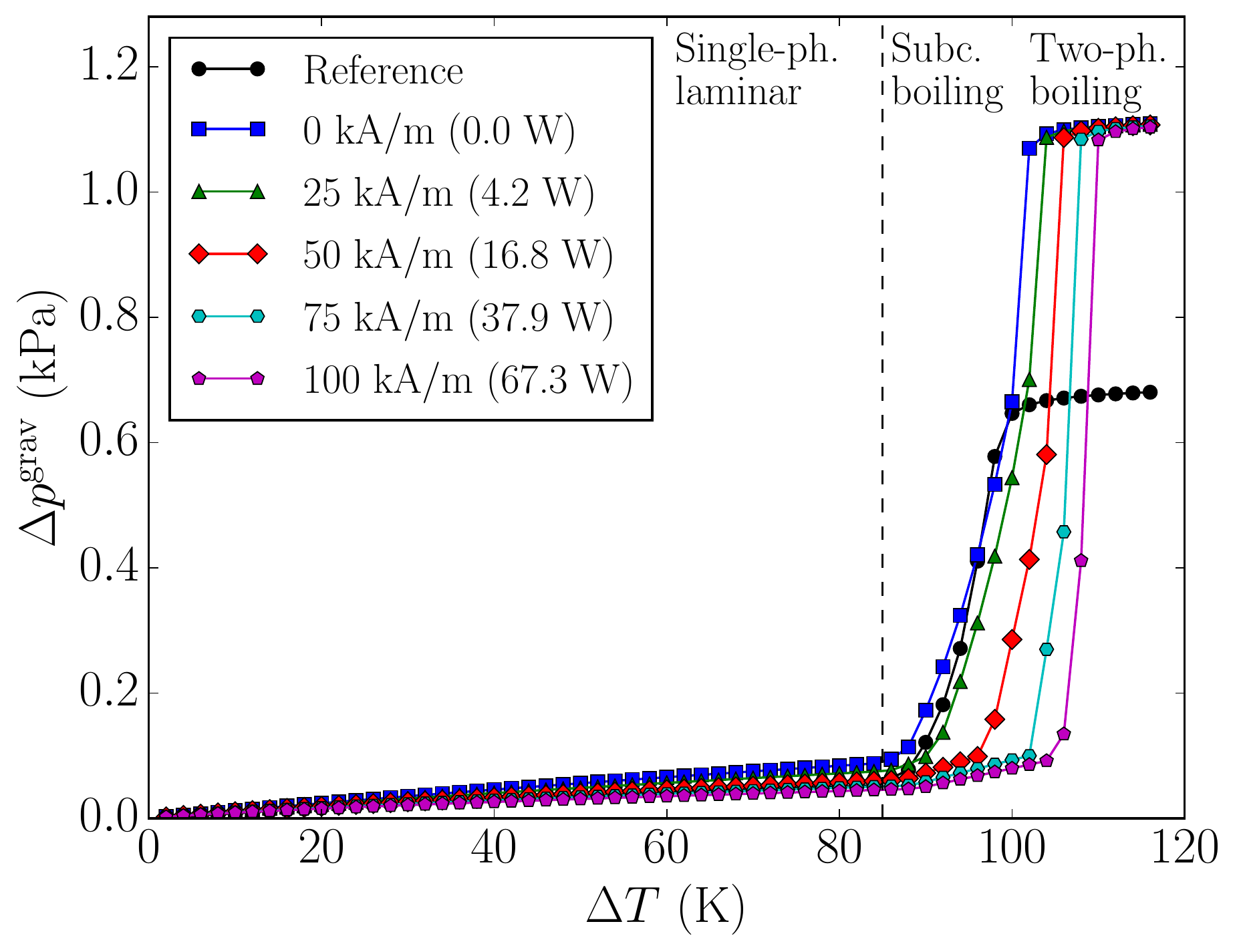}
        \caption{The pressure contribution from natural convection
          calculated from simulation results through \cref{eq:dp_grav_from_sim},
          for varying $\Delta T$ and $H_\text{max}$.
        }
        \label{fig:dpgrav}
\end{figure}

\begin{figure}[tb]
        \centering
        \includegraphics[width=0.48\textwidth]{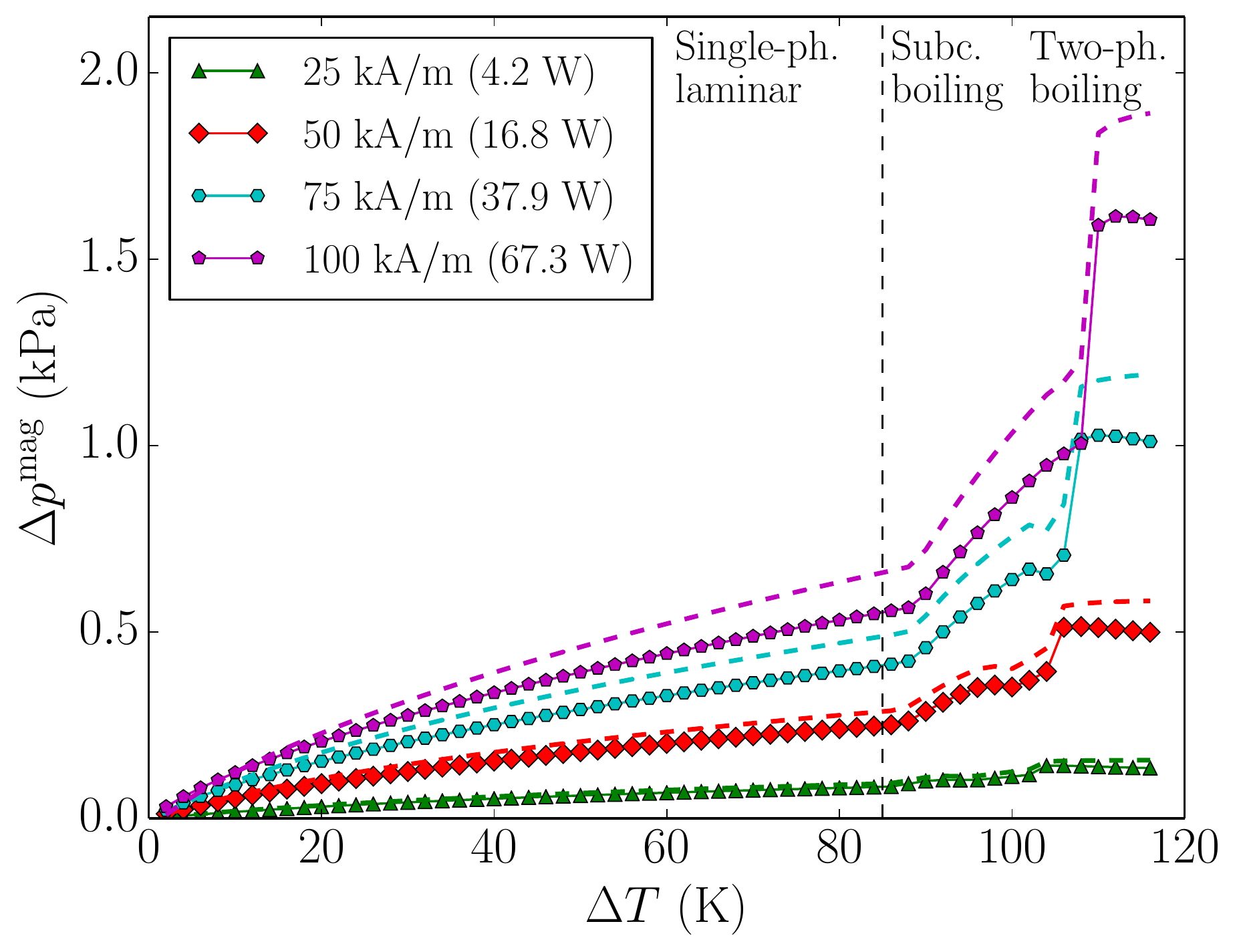}
        \caption{The pressure contribution from the thermomagnetic pump
          calculated from simulation results through \cref{eq:dp_mag_from_sim},
          for varying $\Delta T$ and $H_\text{max}$.
          Above each line is a dashed line showing the approximation found
          by using \cref{eq:dp_approx_from_delta_chi}, where only
          $\Delta \chi$ is taken from the simulation results.}
        \label{fig:dpmag}
\end{figure}

Generally, there are two driving forces for the flow:
the thermomagnetic pump and the natural convection. Each may be
quantified by their net contribution to the pressure-change around the loop.
We may define the pressure contributions as
\begin{equation}
  \Delta p^\text{grav} = \oint f^\text{grav} \dd{x},
  \label{eq:dp_grav_from_sim}
\end{equation}
and
\begin{equation}
  \Delta p^\text{mag} = \oint f^\text{mag} \dd{x},
  \label{eq:dp_mag_from_sim}
\end{equation}
where the gravitational and magnetic force terms are
described in~\cite{Aursand15} and \cref{eq:force_term_mag},
respectively.

The integration is performed around the whole loop in the
counterclockwise flow direction.
Of course, the solution demands that
the total $\Delta p$ around the whole loop is zero.
The cancelling contribution is the friction force.

The validity of the estimate \eqref{eq:dp_approx_from_delta_chi} may be tested
by comparing it to \cref{eq:dp_mag_from_sim} as calculated from a
simulation.
The $\Delta \chi$ to insert in \eqref{eq:dp_approx_from_delta_chi}
is then calculated by evaluating \cref{eq:chi_from_Mp}
at the left and right end of the solenoid from the simulation results.

The calculated values for $\Delta p^\text{grav}$ and $\Delta p^\text{mag}$ for
varying $\Delta T$ and $H_\text{max}$ are presented in
\cref{fig:dpgrav,fig:dpmag}, respectively.
The estimates from \cref{eq:dp_approx_from_delta_chi} are included in
\cref{fig:dpmag} for comparison.

\begin{figure}[t]
        \centering
        \includegraphics[width=0.48\textwidth]{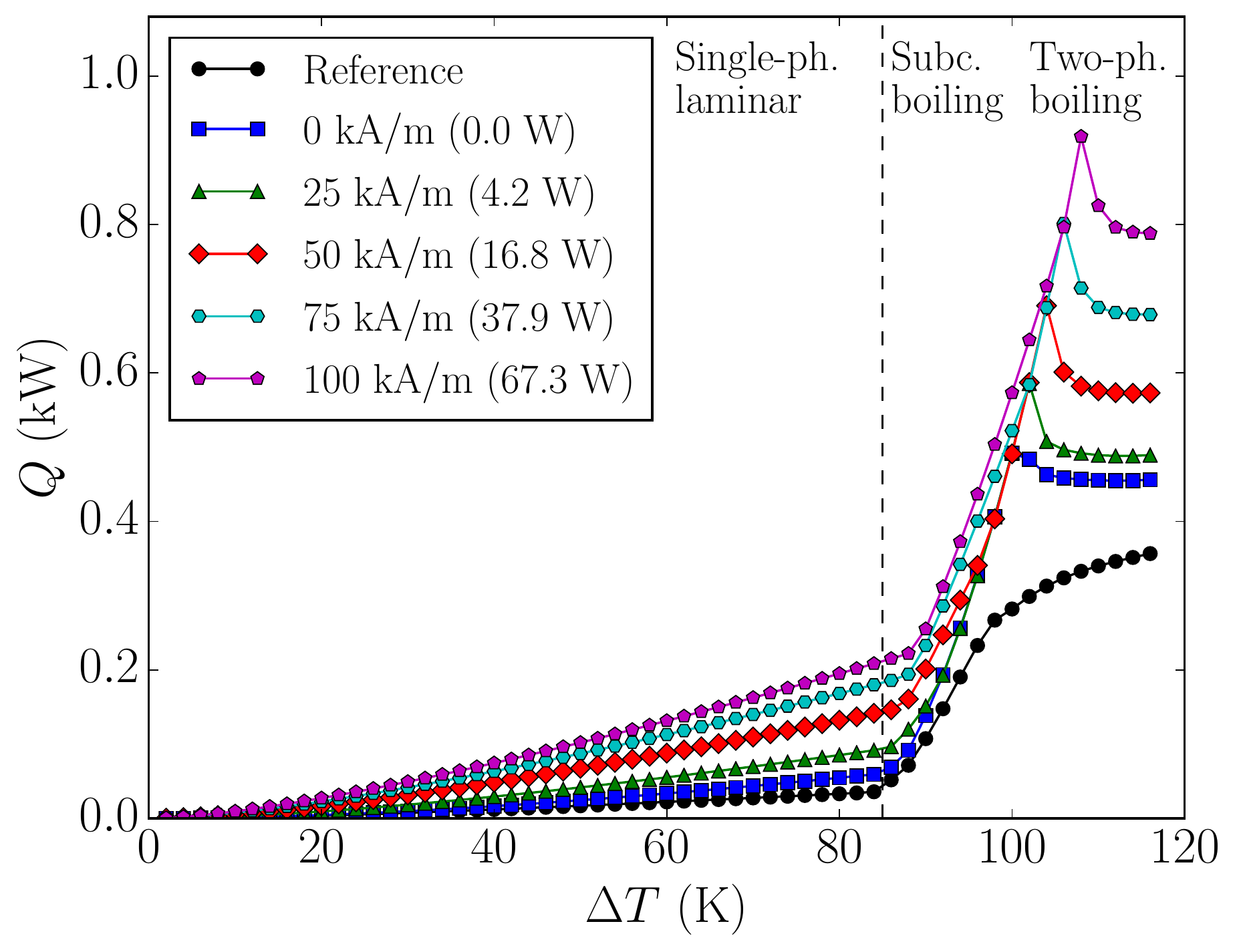}
        \caption{Total rate of heat transfer between heater and cooler,
        for varying $\Delta T$ and $H_\text{max}$, and for
        the reference case with no particles and no field.
        }
        \label{fig:Qtot}
\end{figure}

The total rate of heat transported from the heater to the cooler, denoted as
$Q$~(\si{\watt}), may be used as a performance metric for the whole case. The
calculated values of $Q$ are shown in \cref{fig:Qtot} for varying $\Delta T$
and $H_\text{max}$.

\begin{figure}[t]
        \centering
        \includegraphics[width=0.48\textwidth]{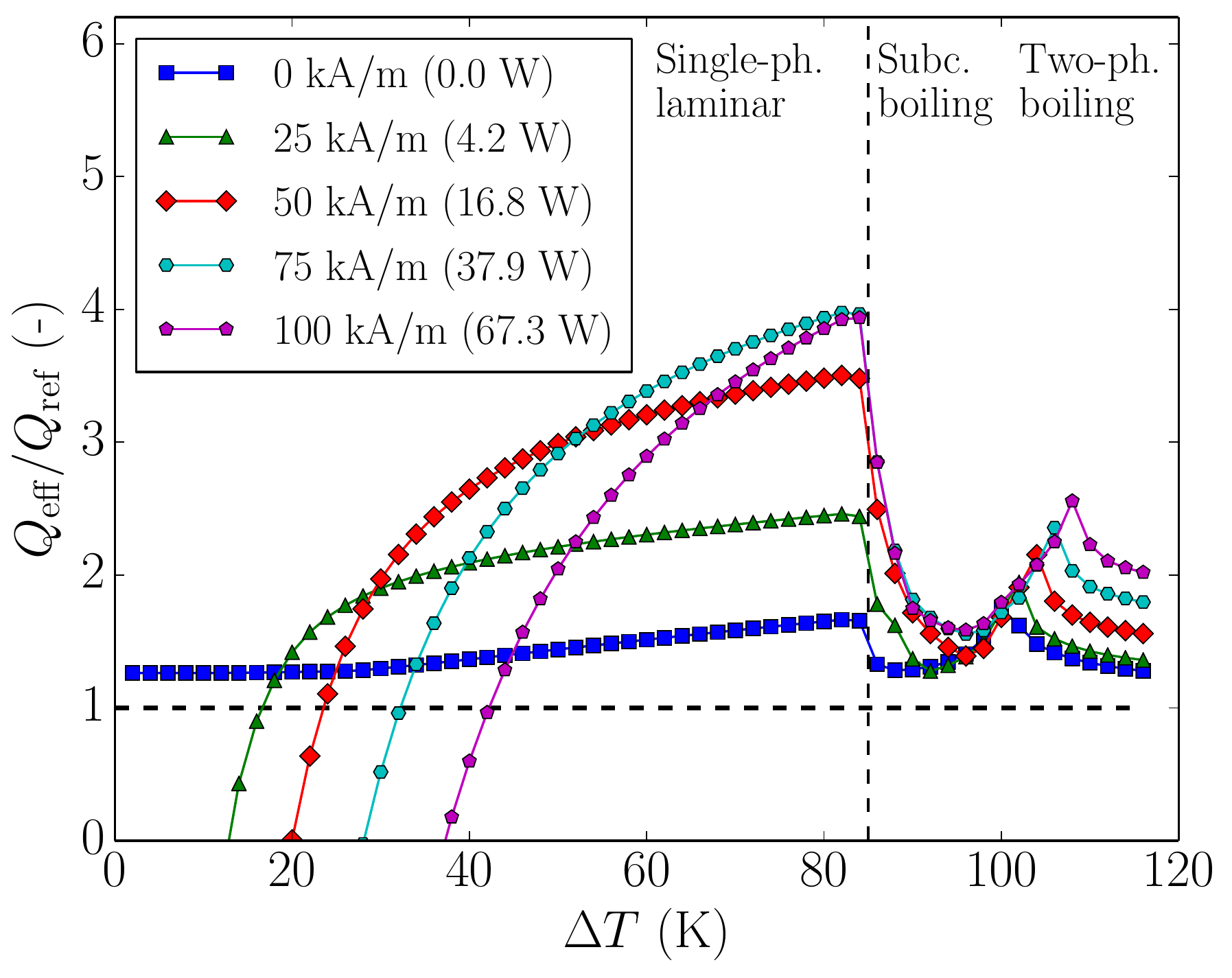}
        \caption{The effective enhancement of cooling power, relative to
        the reference case with no particles and no field, for varying
        $\Delta T$ and $H_\text{max}$. The horizontal dashed line
        shows the limit where the increased heat transfer rate is just enough
        to remove the additional heat added by the solenoid.}
        \label{fig:Qreleff}
\end{figure}

However, for a fair comparison with natural convection, the performance metric
should take into account the power consumption of the solenoid~$P$. The power
consumption is dissipated as heat and adds to the total amount of heat to be
transported away by the fluid. The effective heat that is transported from the
heat source is therefore $Q_\text{eff}$~(\si{\watt}),
\begin{equation}
  Q_\text{eff} \equiv Q - P.
\end{equation}
For the comparison, we calculate the relative improvement or the
\emph{enhancement factor}, that is $Q_\text{eff}/Q_\text{ref}$, where
$Q_\text{ref}$ is the reference result with no particles, that is, pure natural
convection. The effective enhancement is presented in \cref{fig:Qreleff}.

\begin{figure}[ht]
        \centering
        \includegraphics[width=0.48\textwidth]{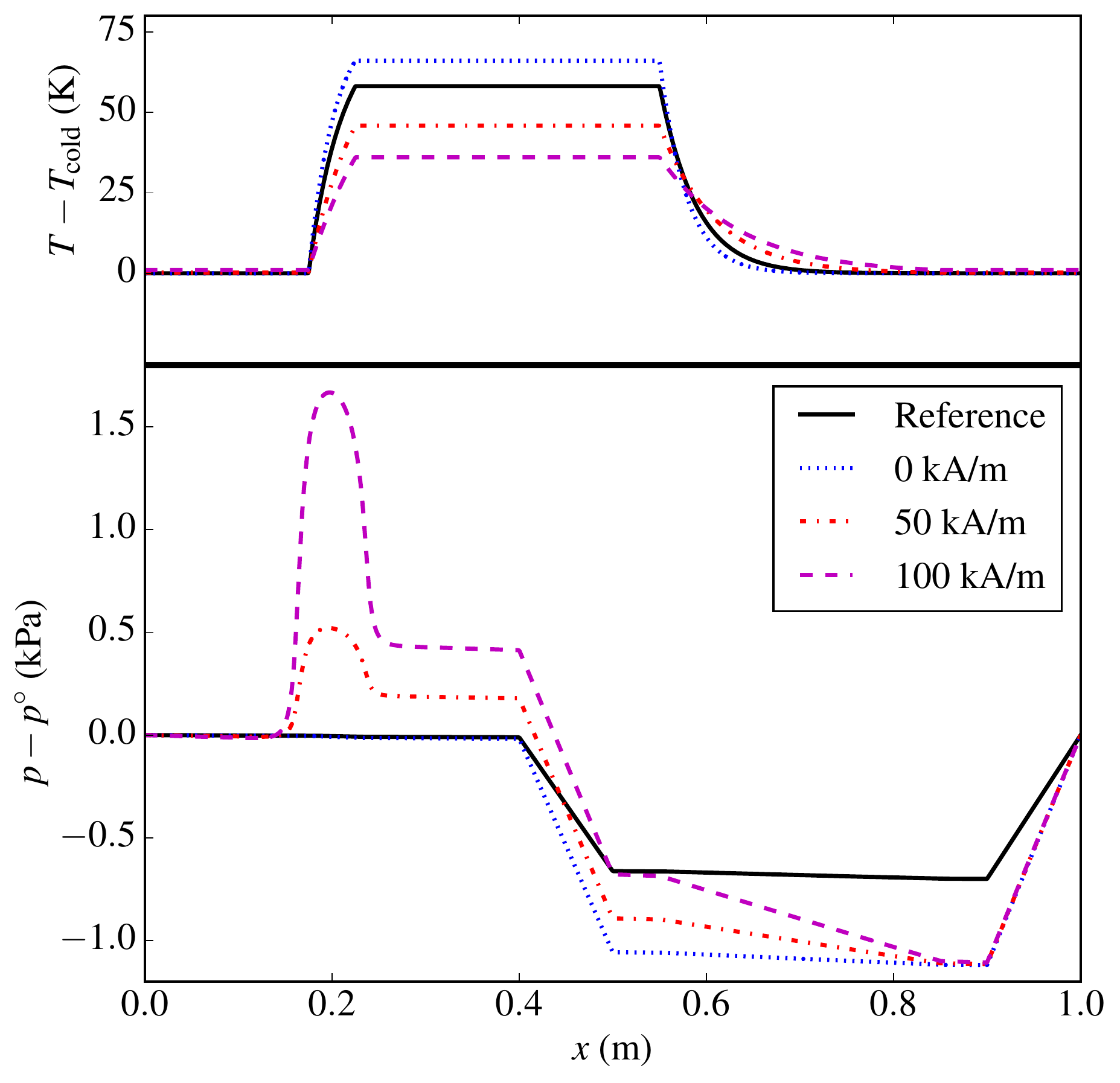}
        \caption{The pressure (bottom), relative to pressure at $x=0$, and
          temperature (top), relative to cooler temperature, plotted 
          against position $x$
        along the rig for a set of cases with $\Delta T = \SI{80}{\kelvin}$.
        The magnetization symmetry breaking can be seen as the pressure jump
        between $x=0.1$ and $x=0.3$, which is approximately equal to $\Delta
        p_\text{mag}$. If there was no symmetry breaking, then there would be no
        pressure difference here.}
        \label{fig:pT_vs_x_80}
\end{figure}

\cref{fig:pT_vs_x_80} shows the pressure and temperature along the rig
at the point of maximum enhancement, i.e.~at about $\Delta
T = \SI{80}{\kelvin}$. At this same point, we also investigated the sensitivity
in the nanofluid heat transfer correlation (HTC). \Cref{tab:sensitivity_Q} shows the
results where the HTC is adjusted with $\pm\SI{20}{\percent}$.

\begin{figure}[htbp]
  \centering
  \includegraphics[width=0.48\textwidth]{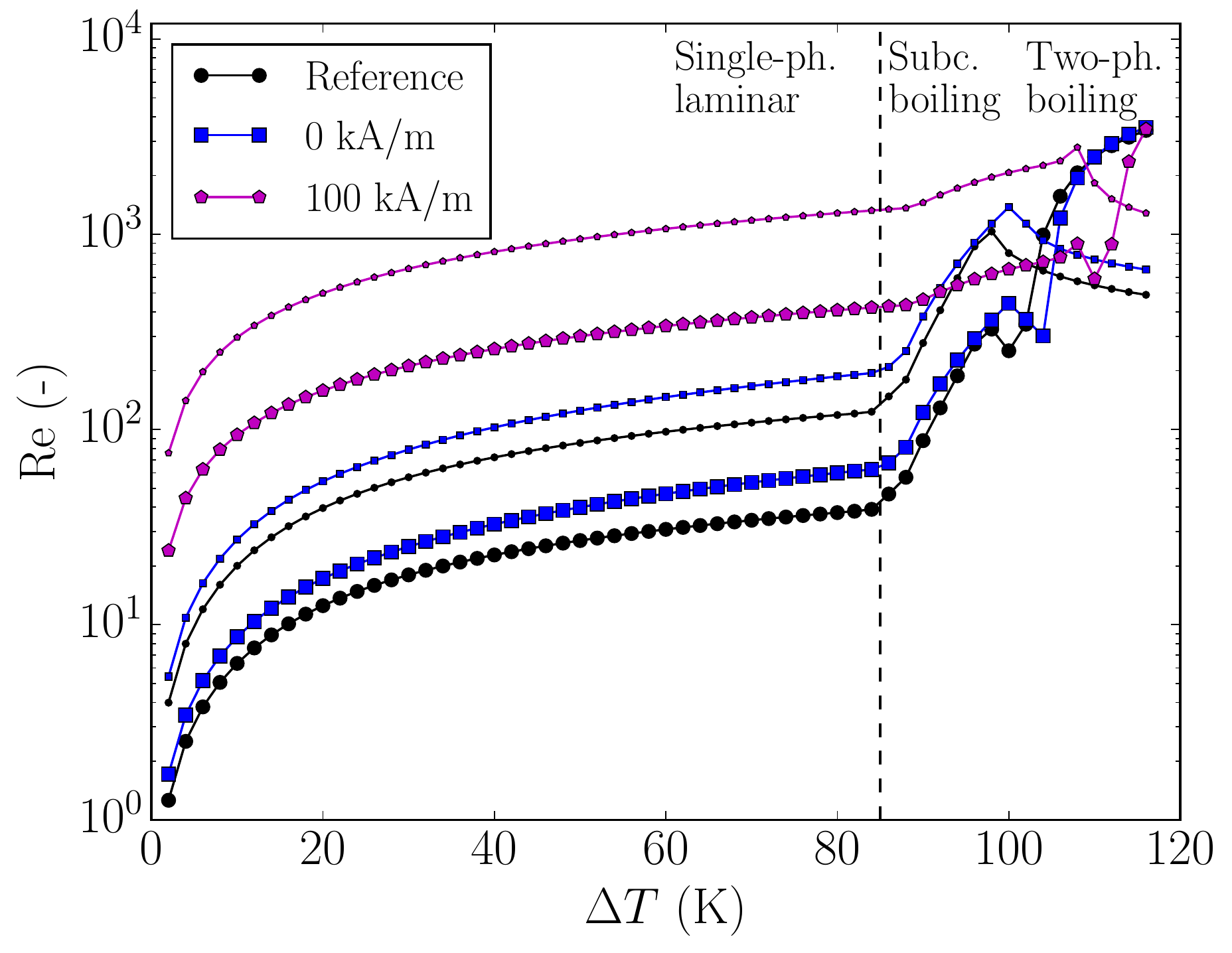}
  \caption{The Reynolds number as a function of temperature difference, as
    measured before the heater (small markers) and at the center of the heater
  (large markers).}
  \label{fig:Re_vs_dT}
\end{figure}

Finally, \cref{fig:Re_vs_dT} shows the Reynolds number in front of the heater
and at the center of the heater as a function of temperature differences, and
for varying magnetic fields.

\begin{table}[htpb]
  \centering
  \caption{Effects on total heat transfer $Q$ (\si{\watt}) by
    $\pm\SI{20}{\percent}$ in the nanofluid heat transfer correlation for the
    $\Delta T = \SI{80}{\kelvin}$ case. Percentages are relative to the first
  row.}
  \label{tab:sensitivity_Q}
  \begin{tabular}{ccrcr}
    \toprule
    & \multicolumn{2}{l}{$H=\SI{0}{\kilo\ampere\per\meter}$}
    & \multicolumn{2}{l}{$H=\SI{100}{\kilo\ampere\per\meter}$} \\
    \midrule
    Orig.   & 54.7 & & 195.1 \\
    $+\SI{20}{\percent}$ & 60.7 & ($+\SI{11}{\percent}$) & 235.5 & ($+\SI{21}{\percent}$) \\
    $-\SI{20}{\percent}$ & 47.2 & ($-\SI{14}{\percent}$) & 153.4 & ($-\SI{21}{\percent}$) \\
    \bottomrule
  \end{tabular}
\end{table}

\section{Discussion}
\label{sec:discussion}

\subsection{Driving forces}

The two driving forces for flow, in terms of their contribution to
pressure increase around the loop, were shown in
\cref{fig:dpgrav,fig:dpmag}. This section will explain and discuss their
features as the two main variables are changed.

The natural convection is driven by a difference in
fluid density at the right and left
ends of the loop. The density difference is caused by
thermal expansion of the liquid due to temperature differences,
as well as the presence of vapor on the right side in the boiling cases.

On both ends of the solenoid, the magnetic force points in towards the
solenoid center.
The key to the thermomagnetic pumping effect is the
magnetization symmetry breaking, which is to reduce the downstream
magnetic force compared to the upstream one, and thus achieve a
positive $\Delta p^\text{mag}$, as seen in \cref{fig:pT_vs_x_80}.
The thermomagnetic pumping is therefore
driven by a decrease of susceptibility $\chi$
across the heater (see \cref{eq:dp_approx_from_delta_chi}),
and as mentioned in \cref{sub:ferrofluid_design}, there are
two mechanisms for this. First, an increase in fluid temperature decreases
the susceptibility of the particle phase, as shown in \cref{fig:mag_curves}.
Second, a decrease in base fluid density increases the flow velocity, and
thus decreases the concentration of particles $\alpha_\partic$ by
``stretching'' the particle distribution.
The effect of the thermomagnetic pump may also be seen in \cref{fig:Re_vs_dT},
which shows that the Reynolds number increases when the magnetic field is
increased.

The results, presented in \cref{fig:dpgrav,fig:dpmag}, reveal three distinct regimes: A single-phase regime, a transition
regime with sub-cooled boiling, and a two-phase regime.
Below approximately $\Delta T = \SI{85}{\kelvin}$, the heater temperature
is below the local bubble temperatures. Here there is no vapor, nor or any kind
of boiling heat transfer.
Still, increasing $\Delta T$ increases the fluid
temperature differences, and thus both driving forces increase.
Note that in this regime, $\Delta p^\text{mag}$ dominates over
$\Delta p^\text{grav}$ at the higher field strengths, and thus the
thermomagnetic pumping is the major driving force.

At higher $\Delta T$, in the transition regime, sub-cooled boiling is active at
the heater.
This means that the heater temperature is above the
bubble temperature, but the average base fluid state is still in the
liquid region. In this regime, local boiling at the walls increase the heat transfer
coefficient at the heater. This increase leads to a rapid enhancement of both driving forces.

In the two-phase regime where the fluid temperature after the heater has passed
the bubble temperature, we enter the saturated boiling heat transfer regime.
While the vapor fraction on the right vertical section increases due to
boiling, $\Delta p^\text{grav}$ increases rapidly with $\Delta T$, as seen in
\cref{fig:dpgrav}. The driving force eventually
reaches a final plateau where most of the hexane is vaporized, in which case there
is no more density difference to be gained.
The height of this plateau is independent of the
magnetic field, since it does not affect the liquid-vapor density
difference. However, it does depend on the presence of particles,
as they affect the mass of the ferroliquid phase.

As seen in \cref{fig:dpmag}, the thermomagnetic pumping force also sees a
rapid increase as two-phase flow starts to occur. 
This is due to the sudden decrease in $\alpha_\partic$ compared to the
heater inlet, which decreases $\chi_\text{R}$, and thus increases
$\Delta \chi$.
This driving force also reaches an optimal plateau as the vapor fraction reaches its
plateau. On this plateau, the symmetry breaking
of the magnetization response is near its optimum, as
$\chi_\text{R} \ll \chi_\text{L}$. The pressure jump $\Delta p^\text{mag}$ is
then approximately given by
\cref{eq:dp_approx_from_delta_chi} with $\chi_\text{L}$ substituted
in place of $\Delta \chi$. We also remark here that the simple calculation of
the pumping force given at the end of \cref{sec:analytical_estimates}, that is,
$\Delta p^\text{mag} \sim \SI{1}{\kilo\pascal}$ for $H_\text{max}
= \SI{100}{\kilo\ampere\per\meter}$, gives a coarse but reasonable estimate.

As opposed to the plateau of $\Delta p^\text{grav}$,
the plateau of $\Delta p^\text{mag}$ depends on $H_\text{max}$.
From \cref{eq:dp_approx_from_delta_chi} we see that it increases as the
square of $H_\text{max}$, or by considering \cref{eq:Hmax_from_Gfactor},
linearly with solenoid power consumption.
Note that the plateau of $\Delta p^\text{mag}$ at the highest field strength
is about \SI{50}{\percent} greater than the plateau of $\Delta p^\text{grav}$. This shows that
the thermomagnetic force can be the major driving force also in the boiling
regime.

From \cref{fig:dpmag}, we see that \cref{eq:dp_approx_from_delta_chi} gives
a reasonable but slightly high estimate for $\Delta p^\text{mag}$.
Given a magnetization model and a field strength,
\cref{eq:dp_approx_from_delta_chi} provides $\Delta p^\text{mag}$ as an
explicit function of $T$ and $\alpha_\partic$ at the heater inlet and outlet,
which could prove very useful for more high-level engineering of
complex systems.
If one wants an even simpler relation, one can assume that
$\chi_\text{R} \ll \chi_\text{L}$ and thus $\Delta \chi = \chi_\text{L}$.
This gives a best-case estimate, which only depends on
$\chi=\alpha_\partic \chi_\partic$ at the inlet.
This corresponds to the final plateaus in \cref{fig:dpmag}.

From \cref{fig:dpgrav},
we see that
$\Delta p^\text{grav}$ is reduced
with increasing magnetic field in the single-phase regime, and the
appearance of vapor is pushed to higher $\Delta T$. This is due to
the added $\Delta p^\text{mag}$ increasing the flow velocity, which suppresses
temperature change when passing the heater and cooler.
However, the added $\Delta p^\text{mag}$ ensures a net increase in
total driving force.





\subsection{Cooling performance enhancements}

The three regimes discussed in the previous section can also be identified in the total heat
transfer rate $Q$,
as seen in \cref{fig:Qtot}. Before the onset of boiling, it increases
smoothly with $\Delta T$. It also increases when particles are added, and
further with the magnetic field. As boiling initiates, an abrupt
increase is seen in a transition phase, before it stabilizes on plateaus
in the saturated boiling regime.

As mentioned, it is important to account for the heat added by the solenoid.  A more fair
performance measure is therefore $Q_\text{eff}$, which may be
interpreted as the maximum power the cooled device may be run at while
still keeping it at the target temperature.
The enhancements of this, relative to the reference case, are seen in
\cref{fig:Qreleff}.

When entering the boiling regime at the heater, $Q$ increases rapidly
for all cases. However, we see a sudden drop in $Q_\text{eff}$, which is due
to the fact that the reference case increases more rapidly. Simply put,
entering boiling heat transfer is a much greater leap from
conventional laminar heat transfer than from
nanofluid-enhanced laminar heat transfer.
When entering saturated boiling and two-phase flow, there is a transition
to turbulent friction in some regions, leading to a decrease in $Q$ in some
cases, but eventually we see a stabilization.

The performance enhancement in a given case can come from two different
effects.
First, there is the \emph{nanofluid effect}, which is the effect of the
fluid having different intrinsic properties compared to if no particles were present.
Second, there is the \emph{thermomagnetic pumping effect}, which is the
effect of adding the magnetic field, which introduces the $\Delta p^\text{mag}$
driving force that was discussed in the previous section.

The most likely $\Delta T$-areas of operation are either the laminar
pure single-phase (no boiling) regime
or well into the saturated boiling regime.
\Cref{tab:enhancements} lists the best possible enhancements in cooling
performance in these regimes,
with the particle amount and field strengths considered here.

\begin{table}[htbp]
  \centering
  \caption{The best performance enhancement factors found compared to the reference
  case, in the single-phase (ferroliquid) and two-phase/boiling
  (ferroliquid+vapor) regimes, from the nanofluid effect and the
   thermomagnetic pumping effect. The first column shows the enhancement seen
   from the addition of nanoparticles alone, while the second column
   shows the additional enhancement seen from thermomagnetic pumping.
   The third column shows the total enhancement with thermomagnetic pumping
   compared to the reference.}
  \label{tab:enhancements}
  \begin{tabular}{lccc}
    \toprule
    & Nanofluid  & Thermomagnetic & Total  \\
    \midrule
    1-phase  & 1.7  & 2.4  & 4.0 \\
    2-phase     & 1.3  & 1.5  & 2.0 \\
    \bottomrule
  \end{tabular}
\end{table}

Note that these results depend on the quite conservatively chosen
particle volume fraction of \SI{10}{\percent}. Higher amounts are possible, and
since both enhancement effects scale with $\alpha_\partic$,
higher enhancements could also be possible. This would be a trade-off
between performance and stability/viscosity concerns.

\subsubsection{Nanofluid effect}
As seen in \cref{fig:Qreleff} and \cref{tab:enhancements},
performance enhancements in the area of \SI{50}{\percent} are within reach, simply
by adding nanoparticles to the base fluid.

We studied which effects in the model that contributed to these performance
enhancements by turning them off one by one, and observed how $Q$
in the new solution differs from the original.
We found that the nanofluid enhancement comes from increases in Nusselt number,
conductivity, and volumetric
heat capacity (see \cref{tab:fluid_properties}). The relative amounts of
each depend on the heat transfer regime.

%
%

\subsubsection{Thermomagnetic pumping effect}

As seen in \cref{fig:Qtot}, $Q$ always increases when the magnetic field
strength increases.
However, when subtracting the solenoid power dissipation,
\cref{fig:Qreleff} reveals that different $H_\text{max}$ are optimal depending
on $\Delta T$.
As seen in \cref{fig:Qreleff} and \cref{tab:enhancements},
the thermomagnetic pumping effect can provide
significant enhancements on top of the nanofluid effect, in both
the single-phase and two-phase boiling regimes.
The most significant enhancements are seen in the single-phase regime,
even though one sees from \cref{fig:pT_vs_x_80} that the symmetry breaking
is far from complete, and much less than in the two-phase boiling cases.
However, in these cases, there are only weak laminar friction forces, and
therefore there is much flow velocity to be gained from $\Delta p^\text{mag}$.

At very low $\Delta T$, the magnetic symmetry breaking, i.e.\ $\Delta \chi$, is
so small that the net thermomagnetic pumping cannot compensate for
the heat dissipated from the solenoid.
When a curve is below the dashed line in \cref{fig:Qreleff}, it is better
to turn the solenoid off.
The magnetic forces may still be large,
but on each side of the solenoid they point inwards with almost equal strength,
giving very little net force.

\subsubsection{Permanent magnets}
\label{ssub:permanent_magnets}

Although the thermomagnetic concept described here is passive with respect to
the lack of moving parts, it may still be argued that the thermomagnetic
solution is not truly passive, since the solenoid requires power input.
However, it is worth noting that the magnetic field strengths used in this study are
well within the range of what is possible with permanent magnets.
Using permanent magnets instead of a solenoid will
eliminate the solenoid power consumption and heat dissipation.
This would reduce power consumption to zero, give higher effective enhancements,
and make it a truly passive concept like natural convection.

As there are no longer any free currents involved,
Maxwell's equations demand that
\begin{equation}
  \int_{-\infty}^\infty H(x) \dd{x} = 0,
\end{equation}
along the central axis,
which prohibits the simple $H(x) \geq 0 $ field-profile possible with
a solenoid. However, even though the $H(x)$ profile from a configuration
of permanent magnets must have a higher number of extrema, both positive and
negative, there is no reason why one cannot obtain a similar symmetry breaking
and net pumping effect.

\subsection{Compactness}
The benefits may also be viewed in terms of compactness of the cooling
system, instead of increased performance given the same size.

One dimension of compactness is the rig
height (originally \SI{10}{\centi\meter}), which is a required
component of the natural convection effect. Adding the
nanofluid- and thermomagnetic effect may allow for the reduction of height
while retaining a performance no less than the reference case.
In the $H_\text{max} = \SI{100}{\kilo\ampere\per\meter}$ case, we see
from \cref{fig:dpgrav,fig:dpmag} that $\Delta p^\text{mag}$ is
larger than $\Delta p^\text{grav}$ in both regimes, which indicates
that one could make the rig completely flat while still retaining some
enhancement compared to the reference. Additional simulations with no
gravity were run to investigate this.
In the single-phase regime, almost all the enhancement remained, since
$\Delta p^\text{mag}$ is so dominant. In the two-phase boiling regime,
an enhancement of 1.5X remained.

Another dimension of compactness is the length of the heat sink
(originally \SI{30}{\centi\meter}), which may be reduced when adding
the nanofluid- and thermomagnetic effect.
It turns out that in both regimes, the heat sink length may be reduced
to about \SI{25}{\percent} of its original size, while retaining the
original performance.


\subsection{Concept limitations}
While the enhancements found seem promising,
it is worth noting that the nature of the concept puts some restrictions on
where it can be used:

\begin{itemize}
  \item Since the pumping force will be from the cold side to the hot side of
    the field source,
    the field source must be placed at the heat source of the loop,
    not the heat sink. Otherwise, the flow will be damped instead of enhanced.
  \item The magnetic field source should stretch across the heat source from
    end to end, to achieve as high a fluid temperature difference as possible
    between the ends.
    The aspect ratio of the heat source must then allow for a solenoid with a
    reasonable field-per-power efficiency, as shown in \cref{fig:map_Gfactor}.
    However, designs using permanent magnets will make the field-per-power
    concern irrelevant.
  \item While we have accounted for the heat dissipation of the solenoid
    through the need for additional heat removal, that does not mean that
    the additional needed electrical energy input is desired.
    However, permanent magnets may also alleviate this, as discussed in
    Sec. \ref{ssub:permanent_magnets}.
\end{itemize}
Note that the above limitations are for the thermomagnetic pumping effect, and
that the nanofluid effect should be more widely applicable.

\subsection{Comments on model limitations}
As in any model, some simplifications are made which could affect the results.
In this section, we will discuss the most relevant of these, and how they may
or may not add uncertainty to the conclusions.

First, regarding the heat transfer coefficients, one could argue that assuming
$\Nu = 3.66$ when the reference fluid is single-phase is a low estimate, as
the flow would not immediately become well developed. This is quantified by the
Graetz number, $\Gz(x^\prime) = D\Re\Pr/x^\prime$, where $D$~(\si{\meter}) is the tube diameter
and $x^\prime$~(\si{\meter}) is the distance from the heater/cooler inlet. It is commonly cited that the
flow is well developed from the point where $\Gr$ drops below $20$. We will here
consider the single-phase cases with the highest $\Re$, $\Delta T = \SI{80}{\kelvin}$.
In the reference case, we see $\Re\approx 40$ in the bundles
(cf.~\cref{fig:Re_vs_dT}),
and $\Gz < 20$ is satisfied
about \SI{40}{\percent} into the heater. This means it is mostly well developed, but we 
underestime the average $\Nu$ to some degree.
In the  $H=\SI{100}{\kilo\ampere\per\meter}$ ferrofluid case, we see $\Re \approx 400$
in the bundles and $\Gr \approx 80$ at the heater outlet. This means that 
the flow does not have time to fully develop. As the Xuan--Li correlation is presumably
also for developed laminar flow, we will here also underestimate the actual $\Nu$.
However, we will underestimate the nanofluid $\Nu$ to a greater degree, and we therefore 
argue that the found enhancements from the nanofluid effect, \cref{tab:enhancements},
are conservative.

Second, there is some evidence that the classical model may underpredict
the nanofluid viscosity enhancement~\cite{Rudyak13}.
If this is so, the actual enhancement
from the nanofluid effect would be lower than the one reported here.

However, neither of the above points can disturb the additional enhancements
predicted from the thermomagnetic pumping effect.
The introduction of thermomagnetic pumping may however introduce an additional
viscosity increase, called the
magnetoviscous effect~\cite{Ilg09}.
This effect is not present in the model, but we argue that it will be
negligible under the conditions of the cases studied here.
First, even in the noninteracting dipole regime, there may be
rotational magnetoviscosity. However,
the flow in the heater bundle is in the regime where it is negligible,
according to~\cite{Rosensweig69}.
Second, interactions between particles may give additional effects, but
it turns out that
these particles at these temperatures are in the weakly interacting regime,
as defined in~\cite{Ilg09},
where thermal energy prevents magnetic chaining/clustering.
Here the viscosity increase appear to be
around \SI{10}{\percent} at worst, and less in the linear magnetization regime.
In any case,
magnetoviscosity is only present at the solenoid, which is a small part of
the loop as a whole, and thus contributes relatively little to the total
friction.

As concluded in the experimental validation of this model~\cite{Aursand15},
the thermomagnetic pumping is very sensitive to the temperature change
across the solenoid, and thus to the HTC model used. \Cref{tab:sensitivity_Q}
shows the sensitivity of the total heat transfer to variations in the HTC at
$\Delta T = \SI{80}{K}$ for the reference case $H = \SI{0}{kA/m}$ and for $H
= \SI{100}{kA/m}$.
The table shows that the relative uncertainty in total heat transfer is of the same order of magnitude as the relative uncertainty in HTC.



\section{Conclusions and further work}
\label{sec:conclusions_and_further_work}

For the simplified cooling system studied here, we may conclude that:
\begin{itemize}
  \item The thermomagnetic driving force is significant compared to
    natural convection, and in the single-phase case it can be
    several times larger.

  \item Using a ferrofluid can provide significant performance enhancements
    for the cooling system:
    \begin{itemize}
      \item The nanofluid effect alone can improve performance
        by a factor 1.7 in the single-phase laminar cases, and 1.3 in the two-phase
        boiling cases.
      \item Adding the thermomagnetic effect with a solenoid can give
        an additional performance enhancement of 2.4 (4.0 total) in
        the single-phase laminar cases, and 1.5 (2.0 total) in the two-phase
        boiling cases, even when taking into account the additional heat
        from ohmic losses in the solenoid.
    \end{itemize}
      \item Using a ferrofluid can provide significant compactness while retaining
        original (or better) cooling performance:
        \begin{itemize}
      \item The thermomagnetic effect can completely replace the natural convection
        driving force, while still retaining a cooling performance enhancement of
        \SI{50}{\percent} or more. This eliminates the required rig height, enabling more
        compact solutions.
      \item Using a ferrofluid can allow reductions of heat sink size to about
        \SI{25}{\percent} of original size while retaining original performance.
    \end{itemize}
\end{itemize}

Additionally, we have:
\begin{itemize}
  \itemsep0em
  \item Derived design rules for optimal field-per-power solenoid,
    as shown in \cref{ssub:maximum_field_strength}.
  \item Derived design rules for optimal ferrofluid for thermomagnetic pumping,
    as shown in \cref{sub:ferrofluid_design}.
  \item Derived and validated a reasonably accurate approximation
    for the thermomagnetic pumping performance, shown in
    \cref{eq:dp_approx_from_delta_chi} and \cref{eq:dp_from_P},
    which may be useful for higher level
    engineering considerations.
\end{itemize}

Note that the performance enhancements found are specific to this case study,
and may vary with size scales and variations in rig design. However, this study
does demonstrate a great practical potential for using thermomagnetically
pumped ferrofluids to enhance heat transfer systems.
Further studies should explore the range of applicability.

The possibility of using permanent magnets in place of solenoids, as
mentioned in \cref{ssub:permanent_magnets}, is very interesting for
applications where additional power consumption is inconvenient.
Similar studies could explore this using the same model, only using a
$H(x)$ calculated from a configuration of permanent magnets.

\section*{Acknowledgements}
The research project is funded by the Blue Sky instrument of SINTEF
Energy research through a Strategic Institute Programme (SIP) by the
national Basic Funding scheme of Norway.



\bibliographystyle{elsarticle-num-names}
\bibliography{literature}

\end{document}